# Odd-even layer-number effect and layer-dependent magnetic phase diagrams in MnBi$_2$Te$_4$

Shiqi Yang[1,2,3,†], Xiaolong Xu[1,†], Yaozheng Zhu[1,†], Ruirui Niu[1], Chunqiang Xu[4,5], Yuxuan Peng[1], Xing Cheng[1], Xionghui Jia[1], Xiaofeng Xu[4], Jianming Lu[1,\*], Yu Ye[1,2,6,\*]

[1]State Key Laboratory for Mesoscopic Physics and Frontiers Science Center for Nano-optoelectronics, School of Physics, Peking University, Beijing 100871, China

[2]Collaborative Innovation Center of Quantum Matter, Beijing 100871, China

[3]Academy for Advanced Interdisciplinary Studies, Peking University, Beijing 100871, China

[4]Department of Applied Physics, Zhejiang University of Technology, Hangzhou 310023, China

[5]School of Physics and Key Laboratory of MEMS of the Ministry of Education, Southeast University, Nanjing 211189, China

[6]Beijing Key Laboratory for Magnetoelectric Materials and Devices, Beijing 100871, China

[†]These authors contributed equally to this work

[\*]Correspondence and request for materials should be addressed to J.L (jmlu@pku.edu.cn) and Y.Y. (email: ye_yu@pku.edu.cn)

**Abstract**

**The intrinsic magnetic layered topological insulator MnBi$_2$Te$_4$ with nontrivial topological properties and magnetic order has become a promising system for exploring exotic quantum phenomena such as quantum anomalous Hall effect. However, the layer-dependent magnetism of MnBi$_2$Te$_4$, which is fundamental and crucial for further exploration of quantum phenomena in this system, remains elusive. Here, we use polar reflective magnetic circular dichroism** **spectroscopy, combined with theoretical calculations, to obtain an in-depth understanding of the layer-dependent magnetic properties in MnBi$_2$Te$_4$. The magnetic behavior of MnBi$_2$Te$_4$ exhibits evident odd-even layer-number effect, i.e. the oscillations of the coercivity of the hysteresis loop (at $\mu_0 H_c$) and the spin-flop transition (at $\mu_0 H_1$), concerning the Zeeman energy and magnetic anisotropy energy. In the even-number septuple layers, an anomalous magnetic hysteresis loop is observed, which is attributed to the thickness-independent surface-related magnetization. Through the linear-chain model, we can clarify the odd-even effect of the spin-flop field and determine the evolution of magnetic states under the external magnetic field. The mean-field method also allows us to trace the experimentally observed magnetic phase diagrams to the magnetic fields, layer numbers and especially, temperature. Overall, by harnessing the unusual layer-dependent magnetic properties, our work paves the way for further study of quantum properties of MnBi$_2$Te$_4$.**

**Introduction**

Recently, the research on the topological quantum materials has aroused tremendous interest and gained more and more attention in condensed matter physics[1-5]. Materials that combine magnetic and topological properties will reveal more exotic states, such as quantum anomalous Hall (QAH) insulators and axion insulators[6-9]. So far, such magnetic topological insulators (TIs) have been obtained by introducing magnetic atoms into TIs or using proximity effects in magnetic and topological materials heterostructures, however, the related exotic effects can only be observed at extremely low temperatures[6,10-12]. The recently discovered layered MnBi$_2$Te$_4$, showing an out-of-plane ferromagnetic coupling within the layer and antiferromagnetic coupling between the adjacent layers (A-type AFM), is found to be an intrinsic magnetic TI with antiferromagnetism[13-18]. The effective combination of antiferromagnetic order and nontrivial topological energy band makes $MnBi_2Te_4$ a promising material to discover novel topological phases and magnetic phase transitions by either controlling its crystal structures or applying magnetic fields[19-22]. Through complicated sample preparation processes, QAH and topological axion states were probed by low-temperature electrical transport measurements in atomically thin flakes of $MnBi_2Te_4$[23,24]. However, comprehensively revealing the magnetic phase transitions of $MnBi_2Te_4$ under varying external magnetic field, temperature, and the number of layers has not been studied yet, which is of great significance for further exploration of the rich topological phenomena under different magnetic phases.

Polar reflective magnetic circular dichroism (RMCD) spectroscopy, which measures the differential absorption of left and right circularly polarized light induced by the out-of-plane magnetization of the sample (parallel to the light propagation), is a non-destructive optical method for measuring and imaging the magnetism of micro-sized flakes[25-27]. Owing to the small size of the laser spot (~2 μm in diameter), the RMCD spectroscopy measurements are less influenced by inhomogeneity of structural (the domain size of the $MnBi_2Te_4$ bulk was measured to be tens of $μm^2$)[28], enabling subtle magnetic phases originating from finite-size effects in few-number (few-$N$) SLs $MnBi_2Te_4$ flakes to be detected. In addition, RMCD measurement does not require a complicated sample preparation process, which reduces fabrication-induced surface damages, and is very suitable for layer-dependent magnetic studies.

In this work, we utilize RMCD measurement (see the setup in Supplementary Fig. 1) to systematically study the magnetic properties of thin flakes, from single septuple layer (SL) to 9 SLs, and 25 SLs $MnBi_2Te_4$ under different applied magnetic fields and temperatures and drew their magnetic phase diagrams. The results show that for a single SL sample, the ferromagnetism is retained, and as the number of layers increases, the Néel temperature $T_N$ of the antiferromagnetic arrangement in adjacent layers increases simultaneously (from 15.2 K of 1 SL to 24.5 K of 25 SLs samples). The magnetic behavior of $MnBi_2Te_4$ exhibits an evident odd-even layer-number oscillation. In the even-number (even-$N$) SLs $MnBi_2Te_4$, an anomalous magnetic hysteresis loop is observed, which is attributed to the thickness-independent surface-related magnetization induced by the possible spontaneous surface collapse and reconstruction. Combining the experimental observation of the spin-flop transition with odd-even effect and theoretical calculations of the linear-chain model, the dependence of spin-flop field on the number of SLs also allows us to accurately determine the interlayer exchange coupling strength (0.68 meV) and the magnetic anisotropy energy (0.21 meV) for the layered MnBi$_2$Te$_4$, thus capture the magnetic phase evolution under the external magnetic field. The mean-field (MF) method allows us to trace MnBi$_2$Te$_4$ flakes' magnetic phase diagrams depending on the applied magnetic field, the number of layers, and temperature. The phase boundaries obtained experimentally agree well with the theoretical calculations, revealing the capability of the MF model to trace the phase transitions in such 2D antiferromagnetic materials. Our experimental and theoretical findings have determined the magnetic phase diagrams of MnBi$_2$Te$_4$ with the number of layers, temperature, and external magnetic field, keeping the promise for future exploration of quantum phenomena in this intrinsic magnetic topological insulator by controlling its magnetic phases.

**Results**

**Layer-dependent ferromagnetism** MnBi$_2$Te$_4$ is a layered ternary tetradymite compound with the space group of $R\bar{3}m$[29], which consists of Te-Bi-Te-Mn-Te-Bi-Te SL stacking through van der Waals (vdWs) force. Below the $T_N$, the spins of Mn$^{2+}$ ions couple ferromagnetically within the SL with an out-of-plane easy axis but have an antiferromagnetic exchange coupling between the adjacent SL (Fig. 1a), showing an A-type AFM order. The room-temperature Raman spectrum of the MnBi$_2$Te$_4$ crystal shows well-resolved $E_g$ (47 cm$^{-1}$), $A_{1g}$ (66 cm$^{-1}$), $E^2_g$ (104 cm$^{-1}$), and $A^2_{1g}$ (139 cm$^{-1}$) Raman modes (see Supplementary Fig. 3a), consistent with previous reports[17,30]. The temperature-independence Raman spectra imply that there is no structure transition in the measured temperature range down to 2 K (see Supplementary Fig. 3b and Fig. 2 for the setup). Atomically thin flakes down to 1 SL were mechanically exfoliated from bulk crystals onto the gold substrates using standard Scotch tape method and subsequently protected by a layer of polymethyl methacrylate (PMMA) (see methods). Fig. 1b and Fig. 3a display typical optical images of 1 SL and few SLs MnBi$_2$Te$_4$ samples showing obvious contrasts in different thicknesses, which are confirmed by atomic force microscopy (AFM) characterizations (see Supplementary Fig. 4 for details). The height line profiles of the 1 SL (Fig. 1c) and the stepped MnBi$_2$Te$_4$ flakes (Fig. 3b) indicate an SL thickness to be ~1.4±0.1 nm, consistent with previous reports[17,22]. Note all the optical and AFM images were obtained after removing PMMA unless otherwise specified.

The magnetic order of few-$N$ SLs MnBi$_2$Te$_4$ was probed by RMCD microscopy as a function of the applied external magnetic field perpendicular to the sample plane. The RMCD signals were collected under a 0.25 μW 633 nm HeNe laser excitation with a spot size of ~ 2 μm (see results under a 532 nm CW laser excitation in Supplementary Fig. 5). Fig. 1d shows the magnetic field dependence of the RMCD signals of 1 SL MnBi$_2$Te$_4$ at a temperature range from 1.6 K to 18 K. The nonzero RMCD signal at zero field and clear hysteresis loop confirm the ferromagnetism of 1 SL MnBi$_2$Te$_4$. As the temperature increases, the hysteresis loop shrinks and disappears at 18 K, indicating a ferromagnetic (FM) to a paramagnetic (PM) phase transition.

To study the layer-dependent magnetism, we investigated the behavior of thin flakes from 1 SL to 9 SLs under a magnetic field sweeping back and forth from +7 T to −7 T at 1.6 K. RMCD signals versus $\mu_0 H$ were shown in Fig. 1e. All measured odd-number (odd-$N$) SLs consistently show an FM behavior with a single hysteresis loop centered at $\mu_0 H = 0$ T (highlighted by the grey shaded area in Fig. 1e), indicating its ferromagnetic feature due to an uncompensated layer. The coercive field $\mu_0 H_c^{odd}$ increases monotonously with the thickness. In an odd-$N$ SLs A-type AFM material, the Zeeman energy at a fixed magnetic field is proportional to the single uncompensated SL magnetization (invariant with the film thickness), while the anisotropy energy adds up with each SL (increases with the film thickness). Thus, a higher magnetic field is required for the Zeeman energy to overcome the anisotropy energy in the thicker odd-$N$ SLs materials, resulting in a larger coercive field $\mu_0 H_c^{odd}$. Surprisingly, we also observed an anomalous magnetic hysteresis loop centered at $\mu_0 H = 0$ T in even-number (even-$N$) MnBi$_2$Te$_4$ SLs, indicating a net magnetization, which is unexpected for an A-type AFM material. We note this anomalous FM response was also observed in Hall resistance measurements of 4 SLs MnBi$_2$Te$_4$ and was attributed to the possible substrate-induced top-bottom surface asymmetry or disorders in the sample[23]. The observed magnetic hysteresis loop is persistent in all the measured even-$N$ SLs MnBi$_2$Te$_4$, and its coercive field, $\mu_0 H_c^{even}$, increases with the film thickness (highlighted by the pink shaded area in Fig. 1e). Coupled with the fact that the $\mu_0 H_c^{even}$ is much larger than the $\mu_0 H_c^{odd}$ regardless of the film thickness, we can conclude that the net magnetization in the even-$N$ SLs samples is much smaller than those in the odd-$N$ SLs samples and it is also not sensitive with the film thickness. Thus, we rule out the possibility of the net magnetization induced by impurities, defects, or disorders, whose magnitude will increase with the film thickness. We attributed the net magnetization in the even-$N$ SLs MnBi$_2$Te$_4$ to the thickness-independent surface-related magnetization. Recent first-principles calculations and STEM results indicated that the abundant intrinsic Mn-Bi and tellurium vacancy in the exfoliated surface would cause a spontaneous surface collapse and reconstruction in few-layer MnBi$_2$Te$_4$, which might be the origin of the surface magnetization[31].

Under larger magnetic fields, both the odd-$N$ (except for the 1 SL) and even-$N$ SL s flakes undergo spin-flop transitions[14,17] and evolve into complete out-of-plane magnetization above the spin-flip transition fields ($\mu_0H_2$)[32,33]. However, for the flakes with $N \geq 4$, the spin-flip fields are too large that beyond the magnitude of the magnetic field we can apply[33]. The spin-flop transitions exhibit strong odd-even layer-number effects. The spin-flop fields ($\mu_0H_1$) in the odd-$N$ flakes are much larger than those in the even-$N$ flakes, and it decreases (slightly increases) with the film thickness in the odd-$N$ (even-$N$) samples.

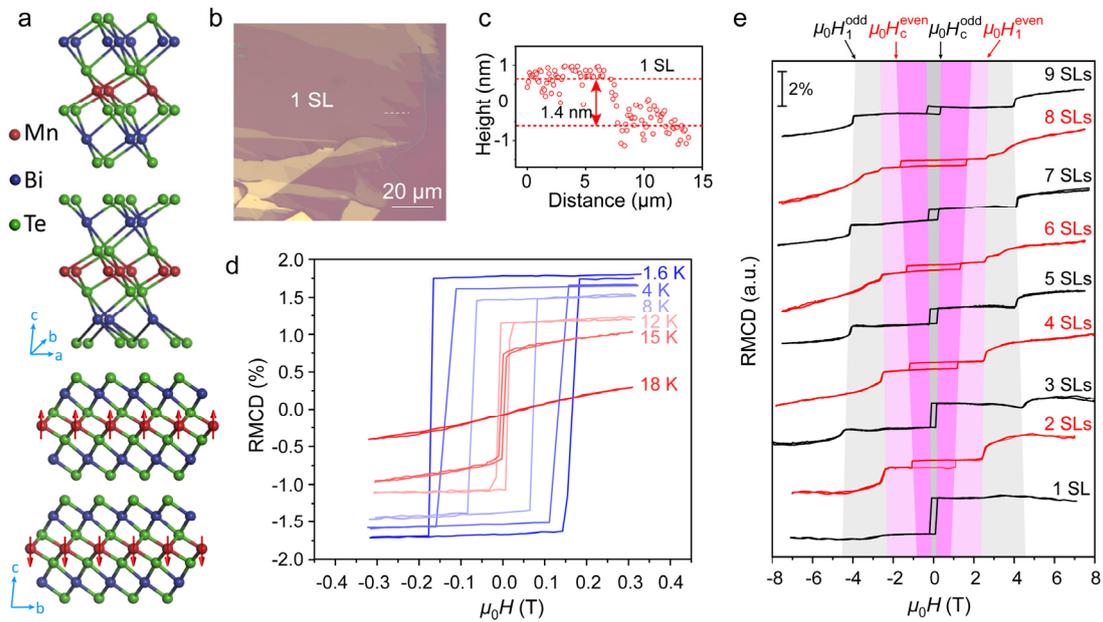

**Fig. 1 Crystal structure and RMCD measurements of 1 SL to 9 SLs MnBi$_2$Te$_4$ flakes. a** Crystal structure of MnBi$_2$Te$_4$. The septuple atomic layers are stacked through vdWs force. The arrows on atoms denote the magnetic moment of each Mn ion. Without an external magnetic field, the neighboring ferromagnetic SLs couple antiferromagnetically with an out-of-plane orientation. **b** The typical optical image of a 1 SL MnBi$_2$Te$_4$ on the Au substrate. **c** Height line profile of the single SL MnBi$_2$Te$_4$ along the white dashed line in **b**. The step height is 1.4 nm, consistent with the thickness of 1 SL MnBi$_2$Te$_4$. **d** Temperature-dependent RMCD measurements of the 1 SL MnBi$_2$Te$_4$. **e** Low-temperature RMCD measurements of MnBi$_2$Te$_4$ flakes (from 1 SL to 9 SLs). The shaded areas highlight the thickness dependences of the low-field spin-flip and high-field spin-flop phase transitions in odd-$N$ and even-$N$ SLs samples.

**Odd-even layer number dependent pin-flop transitions and magnetic state evolutions** For all measured odd-$N$ (3 SLs, 5 SLs, 7 SLs, and 9 SLs) MnBi$_2$Te$_4$ samples, a spin-flop transition at ~ 4 T was observed (Fig. 2a). The spin-flop field of the measured odd-$N$ flakes ($\mu_0 H_1^{odd}$) decreases as the number of layers increases (green circles and corresponding error bars in Fig. 2b). In contrast, the spin-flop transition in the measured even-$N$ (2 SLs, 4 SLs, 6 SLs, and 8 SLs) MnBi$_2$Te$_4$ samples occurs at a much smaller field of ~ 2.5 T (Fig. 2d), and the value of $\mu_0 H_1^{even}$ increases slightly with the number of layers (Fig. 2e). There also exhibit distinguishable second transitions in RMCD measurements of 6 SLs and 8 SLs samples (highlighted by the grey arrows in Fig. 2d), which stem from the sharp coherent spin rotations in a narrow magnetic field range (at $\mu_0 H_s$) after the multi-step spin-flop transitions predicted in the theoretical model (see details in Supplementary Information S5 Part IV). This prominent odd-even layer number effect, with a magnetic phase transition occurring at respective finite fields for odd-$N$ and even-$N$ flakes, is representative and useful to understand the origin in terms of the magnetic phases in layered antiferromagnets MnBi$_2$Te$_4$ under an applied magnetic field.

The magnetic phase transitions can be understood quantitatively using an antiferromagnetic linear-chain model, where the magnetization of each layer is represented by a "macro-spin" coupled to its nearest neighbor layers through the interlayer exchange energy $J$. This simplification is effective when the intralayer ferromagnetic coupling is much stronger than the interlayer antiferromagnetic coupling[34], and it is reasonable to assume uniform magnetization within the single layer at zero temperature. For different layers, the magnetization in the $i$-th layer can be fully described by the angle, $\phi_i$, with the normal direction of the sample. When a perpendicular magnetic field is applied, the average energy per unit cell reads

$$U_N = \mu_0 M_s \left[ \frac{H_J}{2} \sum_{i=1}^{N-1} \cos(\phi_i - \phi_{i+1}) - \frac{H_K}{2} \sum_{i=1}^{N} (\cos\phi_i)^2 - H \sum_{i=1}^{N} \cos\phi_i \right] \quad (1)$$

where $M_s$ is the saturation magnetization per unit cell of a single layer. For the convenience of calculation, the exchange energy and anisotropy energy are expressed in the magnetic field scale, namely, $H_J = (2J)/(\mu_0 M_s)$ and $H_K = K/(\mu_0 M_s)$, where $K \geq 0$ is the easy-axis magnetic anisotropy energy. This model works well for layered antiferromagnetic materials in three regions, namely low-anisotropy region ($H_K/H_J = 0$), mid-anisotropy region ($H_K/H_J = 0.3$) and high-anisotropy region ($H_K/H_J = 0.6$), as shown in Supplementary Fig. 6 (see Supplementary Information S5, Part II for details), which allows us to quantitatively analyze the $N$-dependent magnetic phase transitions observed in the MnBi$_2$Te$_4$ system.

From this model, we first quantitatively explained the evolution of the spin-flop field with the thickness. According to the experimental values of the spin-flop field for each $N$ (from 2 to 9), we perform a standard $\chi^2$-fitting for $H_J$ and $H_K$, and the fitting results obtained are $\mu_0 H_J = 5.10$ T and $\mu_0 H_K = 1.58$ T. The theoretically predicted spin-flop fields agree well with those of the experimentally observed values in both odd-$N$ (Fig. 2b) and even-$N$ samples (Fig. 2e). For the odd-$N$ samples, the magnetization of one uncompensated layer contributes a finite Zeeman energy to the total energy under the external magnetic field at AFM state. Therefore, the spin-flop transition in odd-$N$ samples always occurs at a higher magnetic field than in the even-$N$ samples, where no net magnetization is expected. It is worth note that the $\mu_0 H_1^{odd}$ decreases with the thickness since this effect is originated from the additional Zeeman contribution of one individual uncompensated layer competing against the energetic contribution of all layers[34]. For the even-$N$ samples, there is a small deviation of $\mu_0 H_1^{even}$ between theory and experiment (the experimentally observed field values are overall slightly larger than those of the theoretical prediction), which may confirm the existence of net magnetization in their ground AFM states at zero field, e.g. the surface magnetization discussed above. Utilizing the extracted values of $H_J$ and $H_K$, this model also predicts a multi-step spin-flop transition in 4 SLs, 6 SLs, and 8 SLs samples (see Supplementary Fig. 7). We note that a sharp coherent spin rotation that occurs within a narrow magnetic field range closely follows the final spin-flop transition in the 6 SLs and 8 SLs samples (see Supplementary Fig. 7, $N = 6$ and 8), but not in the 4 SLs sample. Combing the fact that the magnetization changes induced by the multi-step spin-flop transitions (apart from the change at $\mu_0H_1$) are remarkably small, we can conclude that the experimentally observed two-step transitions in 6 SLs and 8 SLs samples originate from the first-step spin-flop transition (at $\mu_0H_1$) and the subsequent sharp coherent spin rotation (at $\mu_0H_s$) that follows the multi-step spin-flop transition.

Furthermore, the evolution of the "macro-spins" in each SL with an external magnetic field unravels the nature of the magnetic phase transitions. For the 3 SLs flake (Fig. 2c), the magnetizations in the adjacent layers remain antiparallel to each other along the z-axis direction until $\mu_0H$ reaches $\mu_0H_1^{odd}$ and undergoes a spin-flop transition to a canting AFM (CAFM) state. The magnetizations of the top and bottom layers experience the same spin-flop and canting processes under an external field. With further increase of the magnetic field, the non-collinear magnetizations eventually reach a fully aligned FM state through coherent rotation at $\mu_0H_2$. For the 2 SLs flake (Fig. 2f), the magnetizations in the two layers remain antiparallel to each other until $\mu_0H$ reaches $\mu_0H_1^{even}$ and experience a spin-flop transition to the CAFM state and eventually realize the FM state. The magnetic state evolutions discussed for the 2 SLs and 3 SLs samples are characteristics for all even-N and odd-N SLs samples. For 4 SLs, 6 SLs and 8 SLs samples, an asymmetric state (e.g. $\phi_1 \neq -\phi_4$, and $\phi_2 \neq -\phi_3$ for $N = 4$) is first formed after the first-step spin-flop transition at $\mu_0H_1^{even}$, and then the symmetric state ($\phi_1 = -\phi_4$, $\phi_2 = -\phi_3$ for $N = 4$) is reached after the final spin-flop transition at $\mu_0H_1'$ (see Supplementary Fig. 8 for details), thus the spin-flop transition is completed through multiple steps. Due to the good agreement between the theoretical and experimental results, we conclude that this model reproduces the layer number-dependent magnetic transitions using parameters of interlayer exchange energy $J = 0.68$ meV and magnetic anisotropy energy $K/2 = 0.21$ meV. We also acquired the RMCD intensity maps including 3 SLs, 4 SLs, and 5 SLs MnBi$_2$Te$_4$ flakes under specific external magnetic fields (see Supplementary Fig. 9), revealing that the evolution of the magnetic states is consistent with the theoretical prediction. Meanwhile, the spatial RMCD intensity maps under different magnetic fields also reveal uniform magnetization across each area (tens of micrometers), indicating the high quality of the samples.

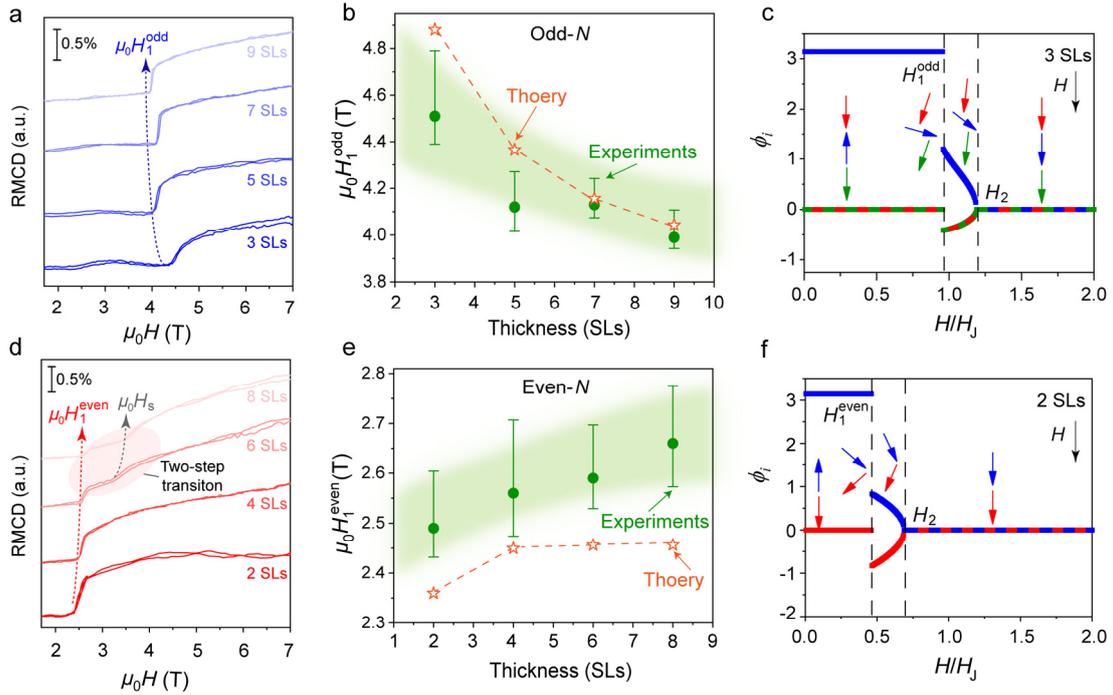

**Fig. 2 Odd-even layer number effect and magnetic state evolutions of thin MnBi$_2$Te$_4$ flakes. a, d** Zoomed-in RMCD measurements of all investigated odd-$N$ samples (**a**) and even-$N$ samples (**d**), showing prominent odd-even layer number effect in MnBi$_2$Te$_4$. In odd-$N$ SLs, all signals exhibit a spin-flop transition at ~ 4 T. The blue dashed arrow represents the spin-flop field in each thickness. In even-$N$ SLs, the spin-flop transition occurs at ~ 2.5 T, which is marked by the red dashed arrow. For 6 SLs and 8 SLs, the RMCD signals exhibit a second transition marked by the grey dashed arrow. **b, e** Spin-flop field versus $N$ for odd-$N$ SLs (**b**) and even-$N$ SLs (**e**) samples. The green circles with error bars denote the experimental results extracted from the RMCD measurements, while orange stars denote calculated theoretical values of spin-flop fields using the parameters of $\mu_0 H_J$ = 5.10 T and $\mu_0 H_K$ = 1.58 T. For the even-$N$ samples, there is a small deviation of $\mu_0 H_1^{even}$ between theory and experiment, which may result from the existence of net magnetization in their ground AFM states at zero field. **c, f** Magnetic state evolutions with the applied magnetic field in 3 SLs (**c**) and 2 SLs (**f**) MnBi$_2$Te$_4$ obtained from the antiferromagnetic linear-chain model. In 3 SLs, the antiferromagnetic state (↑↓↑) is stable until to ~ 4.5 T, where a spin-flop transition occurs, followed by progressively canting until perfect alignment is reached at the spin-flip transition field. In 2 SLs, the AFM state is stable until to ~ 2.5 T, and experience a spin-flop transition to the CAFM state and eventually realize the FM state.

**Thickness – temperature magnetic phase diagram** Then, we discuss the thickness-dependent temperature-driven phase diagram from the AFM phase to the PM phase. The height line profiles (Fig. 3b and Supplementary Fig. 4) helps to clarify the layer number from 1 SL to 9 SLs samples shown in Fig. 3a. Temperature-dependent RMCD measurements of 2 SLs MnBi$_2$Te$_4$ (Fig. 3c) reveal that the $\mu_0H_1$ dwindles as the temperature increases and finally turns into PM at ~20 K. The anomalous magnetic hysteresis loop with a coercive field of ~1.1 T also shrinks with increasing temperatures and disappears at ~20 K. In even-$N$ SLs samples, the value of $T_N$ is estimated by the temperature when the spin-flop transition disappears (see the cases for 4 SLs and 6 SLs samples in Supplementary Fig. 10). As for odd-$N$ SLs samples, we examine their magnetism by focusing on the RMCD intensity of the center magnetic hysteresis loop, due to the magnetization of the uncompensated layer. A clear magnetic hysteresis loop appears in 3 SLs MnBi$_2$Te$_4$ at 20 K (Fig. 3d) but vanishes at 22 K, indicating a magnetic phase transition occurs. The value of $T_N$ can be extracted from the temperature-dependent remnant RMCD signals at $\mu_0H = 0$ T. The temperature-dependent remnant RMCD signals of 1 SL (red), 3 SLs, 5 SLs, and 25 SLs (black) flakes can be well fitted using the critical power-law form $(1-T/T_N)^\beta$, where $T < T_N$, $T_N$, and $\beta$ are two simultaneous fitting parameters (Fig. 3e). The extracted $T_N$ values of 1 SL, 3 SLs, 5 SLs, and 25 SLs flakes are 15.2 K, 22.1 K, 23.4 K, and 24.5 K, respectively, which increase with the sample thickness. The obtained $T_N$ values (open circles for odd-$N$ samples, open triangles for even-$N$ samples) are plotted versus SL-number (Fig. 3f), revealing the phase boundary between the distinct magnetic states—PM and A-type AFM phases. For few-SLs MnBi$_2$Te$_4$ below 10 SLs, the $T_N$ drops from the value of 24.5 K for 25 SLs to 15.2 K for 1 SL sample. We ascribe the suppression of $T_N$ to the increase in thermal fluctuations as the sample approaches the 2D limit.

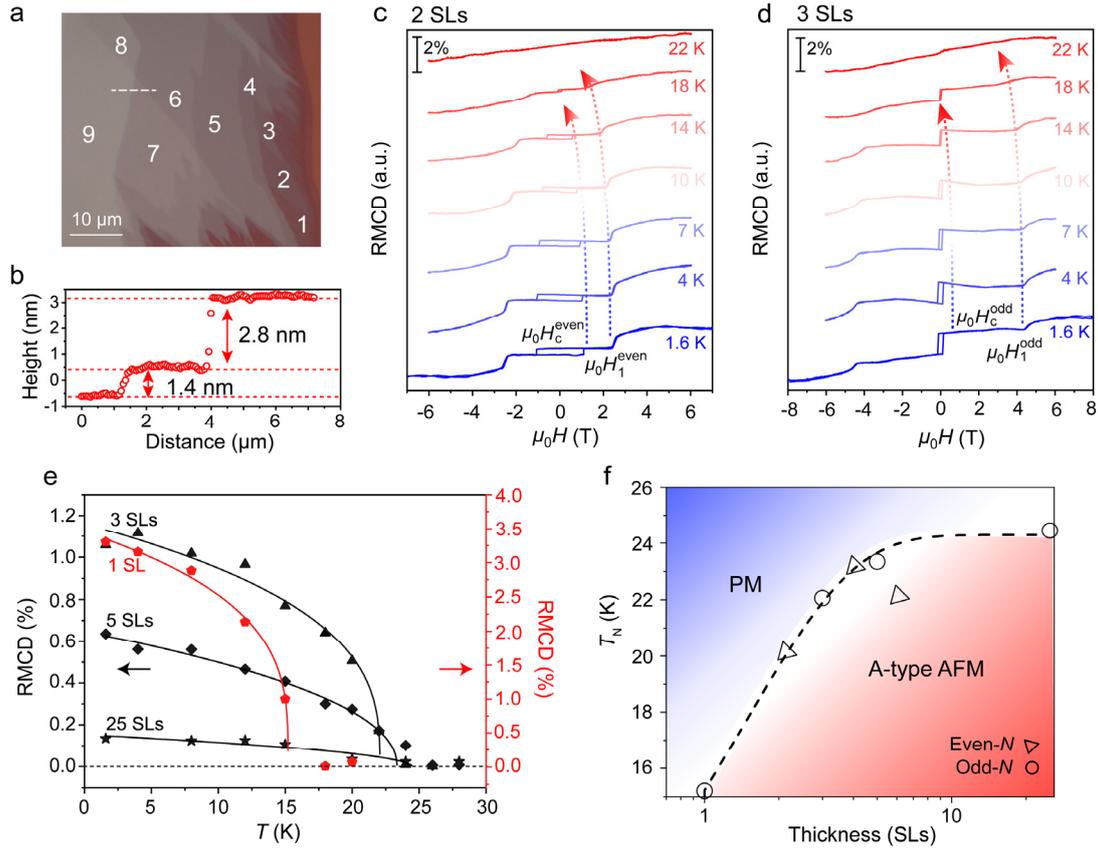

**Fig. 3 Temperature-dependent RMCD measurements and phase diagram of MnBi$_2$Te$_4$ with different thicknesses. a** Optical image of the stepped MnBi$_2$Te$_4$ flakes with 1 SL to 9 SLs. **b** Line height profile of the MnBi$_2$Te$_4$ flakes along with the white dashed line in **a**. **c** RMCD sweeps for the 2 SLs sample at a temperature range that passes through its $T_N$. The spin-flop field ($\mu_0 H_1^{flop}$) decreases as the temperature increases and eventually vanishes at about 20 K. **d** RMCD sweeps for the 3 SLs sample at a temperature range that passes through its $T_N$. The coercive field and remnant RMCD signal of the magnetic loop decrease as the temperature increase, and both eventually vanish at about 22 K. **e** Remnant RMCD signal as a function of temperature for the selected few-$N$ SLs flakes (1 SL, 3 SLs, 5 SLs, and 25 SLs). The solid lines are least-squares criticality fits with the form of $(1-T/T_N)^\beta$ and the black dotted line represents zero RMCD signal. **f** Layer number-temperature phase diagram of the MnBi$_2$Te$_4$ flakes. PM denotes the region where the flake is paramagnetic, A-type AFM denotes the region where adjacent ferromagnetic SLs couple antiferromagnetically with each other.

**Temperature – field phase diagrams** In the above linear chain model, only the ground state is considered, which corresponds to zero temperature. However, "macro-spin" approximation will no longer hold strictly at finite temperatures, so we propose a more precise energy expression (see Eq. (6) in methods), which includes the energy from each spin site and its interactions with every other site. By utilizing the mean-field (MF) method for intralayer interactions to simplify the model, the spin sites are "decoupled" and we can choose one representative spin in each layer to get the $N$-moment energy, which can be written as,

$$U_N^{\mathrm{MF},1} = \sum_{i=1}^{N-1} J \frac{\langle \vec{M}_i \rangle \cdot \vec{M}_{i+1} + \vec{M}_i \cdot \langle \vec{M}_{i+1} \rangle}{M_s^2} + \sum_{i=1}^{N} J^{\parallel} \frac{\vec{M}_i \cdot \langle \vec{M}_i \rangle}{M_s^2} - \frac{K}{2} \sum_{i=1}^{N} \left( \frac{\vec{M}_i \cdot \hat{z}}{M_s} \right)^2 \\ - \mu_0 \vec{H} \cdot \sum_{i=1}^{N} \vec{M}_i - \sum_{i=1}^{N-1} J \frac{\langle \vec{M}_i \rangle \cdot \langle \vec{M}_{i+1} \rangle}{M_s^2} - \frac{1}{2} \sum_{i=1}^{N} J^{\parallel} \frac{\langle \vec{M}_i \rangle \cdot \langle \vec{M}_i \rangle}{M_s^2} \qquad (2)$$

where $J$ represents effective interlayer interaction, $J^{\parallel}$ represents effective intralayer interaction (see supplementary Eq. (14)), and $K$ represents the magnetic anisotropy. Previous extracted values of $J$ and $K$ in the linear chain model can be directly used in this model. Combining this MF method with self-consistent conditions, the magnetization under an external magnetic field that varies with temperature can be obtained. It is worth noting that this model is equivalent to the linear chain model at zero temperature (see Supplementary information S8. Part II for details).

Using this method, we obtained the temperature – field phase diagrams of 2 SLs to 6 SLs samples (Fig. 4, and see more in Supplementary Fig. 11). From temperature – field ($T - \mu_0H$) phase diagrams, it is clear to see the coincidence of the phase boundaries of A-type AFM/CAFM/FM between theoretical predictions (white circles and triangles in Fig. 4) and experimental behaviors (grey spheres and triangles with error bars) up to 18 K for $N = 2$ and 22 K for $N \geq 3$ samples. The spin-flip fields in most measuring results are hard to distinguish experimentally, due to the smooth transition process at $\mu_0H_2$ at finite temperatures. As the temperature approaches the Néel temperature, the theoretical prediction gradually deviates from the experimental data, because fluctuations of two-dimensional magnetic systems become more dominant in this region, which is ignored in the MF method. In addition, as the temperature increases, the spin-flop takes place with a smoother transition as we observed experimentally, and the surrounding hysteresis loop narrows (eventually disappears) in the *M – H* curve at higher temperatures (see Supplementary Fig. 12). All in all, this model predicts the phase diagram very well below the Néel temperature, showing its capability to trace the properties of 2D antiferromagnetic materials.

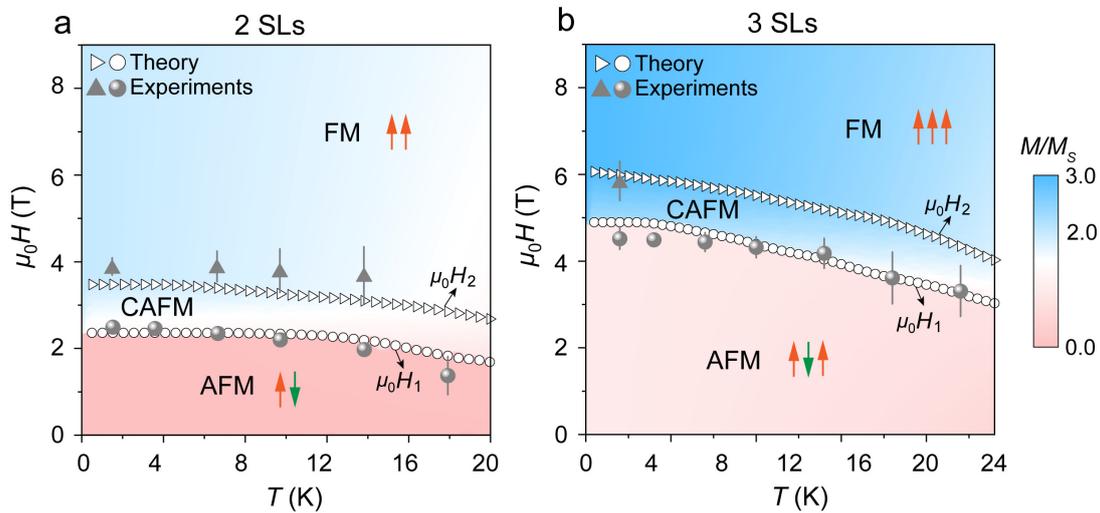

**Fig. 4 Temperature – field phase diagrams of 2 SLs and 3 SLs MnBi$_2$Te$_4$. a** Temperature – field phase diagram of 2 SLs MnBi$_2$Te$_4$ determined by MF methods, which is in coincidence with the RMCD measurements. The white circles and triangles represent the calculated spin-flop field $\mu_0H_1^{flop}$ and spin-flip field $\mu_0H_2$, respectively, at various temperatures, showing the boundaries of the A-type AFM/CAFM phase and CAFM/FM phase. The experimental data points are represented using grey spheres and triangles with corresponding error bars. **b** Temperature –field phase diagram of 3 SLs MnBi$_2$Te$_4$, showing the phase boundaries between uncompensated A-type AFM/CAFM phase and CAFM/FM phase. The scale bar with color from pink to blue in both graphs shows the magnetization values of $M/M_s$ from zero to 3.0. The inset arrows denote the spin orientation of each SL by orange (↑) and green (↓).

## Discussion

In summary, we examined layer-dependent magnetism in atomically thin intrinsic magnetic TI MnBi$_2$Te$_4$ flakes with varying temperature and applied magnetic field using RMCD spectroscopy. An evident odd-even layer-number effect was observed in thin MnBi$_2$Te$_4$ flakes, i.e. the oscillations of the coercivity of the hysteresis loop (at $\mu_0H_c$) and the spin-flop transition (at $\mu_0H_1$). The observed anomalous magnetic hysteresis loop in AFM even-$N$ SLs samples was attributed to surface-induced magnetism, but nevertheless, the origin still needs further study, for instance, its relationship with surface topological states and surface structure reconstruction. The antiferromagnetic linear-chain model provides an excellent quantitative understanding of the experimental observed odd-even layer-number effect in spin-flop transition field oscillation and the two-step transitions in even-$N$ SLs samples with $N \geq 6$, and also captures the evolution of the magnetic states in MnBi$_2$Te$_4$ as a function of the magnetic field and number of septuple layers. Based on the MF approach, the temperature-dependent theoretical analyses show the capability to trace the experimentally determined phase diagrams of the few-$N$ SLs A-type AFM MnBi$_2$Te$_4$ in the mid-anisotropy region. The investigation of the magnetic state evolution with external magnetic field and temperature helps unravel the material's complex magnetic structures and would provide a fundamental understanding for further studying the related quantum states under diverse magnetic phases. This work opens more opportunities for further studying the quantum phenomenon of magnetic TIs and, plausibly, heterostructures integrating MnBi$_2$Te$_4$ with other 2D materials that are endowed with more exotic properties in condensed matter physics.

## Methods

**Crystal growth** Single crystals of MnBi$_2$Te$_4$ were fabricated via a self-flux method. Precisely weighed high-purity manganese powder, bismuth lumps, and tellurium shots were mixed with a molar ratio of 1:10:16, filled into an alumina crucible, and then sealed in a quartz tube under the vacuum better than 5 mtorr. The tube was heated to 900 °C at a rate of 10 K/min and kept at this temperature for one day to ensure complete mixing. The tube was then slowly cooled to 600 °C and then quenched with cold water. The shiny, plate-shaped MnBi$_2$Te$_4$ single crystals with a length of several millimeters were obtained.

**Sample preparation** Thin MnBi$_2$Te$_4$ flakes with different thicknesses were first mechanically exfoliated on a polydimethylsiloxane (PDMS) substrate, and then transferred onto a gold film evaporated on top of a 285 nm SiO$_2$/Si substrate, as reported in previous work in details[35]. Then, a layer of PMMA was spin-coated on the MnBi$_2$Te$_4$ flakes for protection.

**RMCD measurements** The RMCD measurements were performed based on the Attocube closed-cycle cryostat (attoDRY2100) down to 1.6 K and up to 9 T in the out-of-plane direction. The sample was moved by an $x$–$y$–$z$ piezo stage (Piezo Positioning Electronic ANC300). A 633 nm HeNe laser with linear polarization was coupled into the system using free-space optics (see Supplementary Fig. 1 for details). The linearly polarized light was modulated between left and right circular polarization by a photoelastic modulator (PEM) at 50.052 kHz, and a chopped at a frequency of 789 Hz. Using a high numerical aperture (0.82) objective, a Gaussian beam with a 2 μm in diameter spot size was focused onto the sample surface. The reflected light was also collected by the free-space optics and detected by a photomultiplier tube. Due to the polar magneto-optic effect, the magnetization information was detected by the RMCD signal determined by the ratio of the a.c. component at 50.052 kHz and the a.c. component at 789 Hz (both were measured by a two-channel lock-in amplifier Zurich HF2LI).

**Raman spectroscopy** Raman spectra of thick MnBi$_2$Te$_4$ flake were obtained using the WITec alpha300 confocal innovation system at room temperature. A 532 nm laser was focused by a 50× (0.55 NA, Zeiss) objective onto the sample and the resultant Raman signals were detected using a spectrometer with a 1800g/mm grating coupled with a charged coupled device (CCD). The temperature-dependent low-frequency Raman spectra were obtained using free-space optics base on the Attocube closed-cycle cryostat (attoDRY2100). The detailed optical setup is represented in Supplementary Fig. 3.

**Antiferromagnetic linear-chain model** The core of the antiferromagnetic linear-chain model is to represent the spin moment in a single layer by one equivalent spin, which is coupled antiferromagnetically to its neighboring equivalent spins. At zero temperature, this simplification is valid because the intralayer exchange interactions are much stronger than the interlayer ones, and the ferromagnetic intralayer interaction ensures a uniform magnetization within a single layer for the ground state. Denote the magnetization per unit cell in the $i$-th layer as $\vec{M}_i$, the average energy per unit cell reads

$$U_N = J \sum_{i=1}^{N-1} \frac{\vec{M}_i \cdot \vec{M}_{i+1}}{M_s^2} - \frac{K}{2} \sum_{i=1}^{N} \left(\frac{\vec{M}_i \cdot \hat{z}}{M_s}\right)^2 - \mu_0 \vec{H} \cdot \sum_{i=1}^{N} \vec{M}_i \qquad (3)$$

where $J$ is the interlayer antiferromagnetic coupling, $M_s$ is the saturation magnetization per unit cell of a single layer, $K > 0$ is the easy-axis anisotropy energy, and $H$ denotes the applied magnetic field. Here, the anisotropy includes both magnetocrystalline anisotropy ($K_{mc}$, from spin-orbit coupling in the material) and the shape anisotropy ($K_{sh}$, associated with magnetostatic interactions). Actually $K_{sh} = -\mu_0 M_s^2/V$, where $V$ is the volume of the unit cell. However, $K_{mc}$ is relatively large in MnBi2Te4, and $K = K_{mc} + K_{sh}$ is positive. At zero temperature, we can express that $\vec{M}_i = M_s \hat{e}_i$, where $\hat{e}_i$ is a unit vector. Then the magnetic energy reads

$$U_N = J \sum_{i=1}^{N-1} \hat{e}_i \cdot \hat{e}_{i+1} - \frac{K}{2} \sum_{i=1}^{N} (\hat{e}_i \cdot \hat{z})^2 - \mu_0 M_s \vec{H} \cdot \sum_{i=1}^{N} \hat{e}_i \qquad (4)$$

In our experiment, $\vec{H}$ is along the $z$-axis (out-of-plane). To minimize the energy $U_N$, all $\hat{e}_i$ must be on the same plane (perpendicular to the sample plane). Taking this plane as the $xz$-plane, the magnetization can be expressed as $\vec{M}_i = M_s(\sin\phi_i, 0, \cos\phi_i)$, where $\phi_i$ is the angle between the magnetization in the $i$-th layer and $z$-axis. In terms of $\phi_i$, the magnetic energy reads

$$U_N = J \sum_{i=1}^{N-1} \cos(\phi_i - \phi_{i+1}) - \frac{K}{2} \sum_{i=1}^{N} (\cos\phi_i)^2 - \mu_0 M_s H \sum_{i=1}^{N} \cos\phi_i \qquad (5)$$

By defining the magnetic-field-scale parameters $H_J = (2J)/(\mu_0 M_s)$ and $H_K = K/(\mu_0 M_s)$, then Eq. (1) in the main text can be obtained. See Supplementary Information S5 Part I and Part III for the details about numerically solving the model.

**The mean-field method at finite temperatures** As for the finite temperature, the assumption of uniform magnetization within each layer no longer holds in our system. To propose a more accurate model, we start from the complete magnetic energy of an $N$-layer system

$$U_N = \sum_{i=1}^{N-1} \sum_{a,b} J_{a,b} \frac{\vec{M}_{i,a} \cdot \vec{M}_{i+1,b}}{M_s^2} + \frac{1}{2} \sum_{i=1}^{N} \sum_{a,b} J^{\parallel}_{a,b} \frac{\vec{M}_{i,a} \cdot \vec{M}_{i,b}}{M_s^2} - \frac{K}{2} \sum_{i=1}^{N} \sum_{a} \left(\frac{\vec{M}_{i,a} \cdot \hat{z}}{M_s}\right)^2$$
$$- \mu_0 \vec{H} \cdot \sum_{i=1}^{N} \sum_{a} \vec{M}_{i,a} \qquad (6)$$

where $J_{a,b}$ stands for the interlayer interactions (only the interaction between the atoms in the nearest neighbor layers are accounted), $J_{a,b}^{\parallel}$ stands for the intralayer interactions and $K$ denotes the magnetic anisotropy energy. After applying an MF approximation to the intralayer interactions, the spins at different sites are "decoupled", thus we can take one representative spin per layer to write the corresponding $N$-moment energy as Eq. (2) in the main text (see Supplementary Information S8 Part I for mathematical derivations).

Under the MF approximation, the statistical average value of the magnetic moment in the $i$-th layer is given by

$$\langle \vec{M}_i \rangle = \frac{\int \vec{M}_i \exp\left(-\frac{U_N^{\mathrm{MF},1}}{kT}\right) \mathrm{d}[\theta, \chi]}{\int \exp\left(-\frac{U_N^{\mathrm{MF},1}}{kT}\right) \mathrm{d}[\theta, \chi]} \tag{7}$$

which acts as a self-consistent condition for the energy shown in Eq. (2) in the main text. The status is fully described by $2N$ coordinates $[\theta, \chi] = (\theta_1, \cdots, \theta_N, \chi_1, \cdots, \chi_N)$, then the magnetic moment is defined as $\vec{M}_i = M_s(\sin\theta_i \cos\chi_i,\ \sin\theta_i \sin\chi_i,\ \cos\theta_i)$, with $0 \leq \theta_i \leq \pi, 0 \leq \chi_i < 2\pi$ (like the polar angle and azimuthal angle in a spherical coordinate system). Correspondingly,

$$\mathrm{d}[\theta, \chi] = \prod_{i=1}^{N} \sin\theta_i \mathrm{d}\theta_i \mathrm{d}\chi_i \tag{8}$$

For small $N$, conventional methods are capable of calculating the integral Eq. (7). However, when the integral dimension is larger than 5, Monte Carlo integration methods must be employed. One Monte Carlo step represents a random change of the spin direction, and the energy difference and temperature determine whether the change is accepted. Iterations are performed to meet the self-consistent condition Eq. (7) with the energy Eq. (2) in the main text. At least 512 iteration steps are performed before obtaining the result, ensuring that the self-consistent condition Eq. (7) is verified, through the criterion that the deviation of Eq. (7) among the final 16 iteration steps is less than $0.01M_s$. This also ensures that the iteration does converge. In each iteration step, the thermal equilibrium is reached after 2048 Monte Carlo steps, which is sufficient because the first magnetic state is chosen according to the result of the past iteration step and is thus close to the equilibrium point. In the final iteration steps, the average of at least 65536 Monte Carlo steps are used to determine the observable quantities and the overall result for $\langle \vec{M}_i \rangle$ is thus given by the average of the final 16 iteration steps.

There might be more than one locally stable solutions to Eq. (7), but only one of them will finally be obtained after the iteration. Therefore, to ensure the repetitiveness of this approach, we change the magnetic field from very high (for example, 12 T, when all the magnetic moments are aligned in the same direction) to zero and then back to 12 T, with a step interval smaller than 0.1 T. Within one step, the magnetic state is initialized with the final result of the previous step, ensuring that the states evolve continuously as the magnetic field changes unless one state becomes unstable at a critical magnetic field. Also, to remove the complicated hysteresis loops in the phase diagrams, only one state from the down-sweeping and up-sweeping is kept in the phase diagrams, chosen by the method mentioned in Supplementary Information S8 Part III.

**Acknowledgments**

This work was supported by the National Key R&D Program of China (Grant. Nos. 2018YFA0306900, 2017YFA0206301 and 2019YFA0307800), the National Natural Science Foundation of China (Nos. 61521004, 11974061, and 11974027), and Beijing Natural Science Foundation (Grant. No. Z190011).

**Author contributions**

Y.Y. conceived the project. S.Y. and X.X. performed the measurements. Y.Z. performed the theoretical models and calculations. S.Y., X.X, Y.Z. and Y.Y. analyzed the data and wrote the manuscript. R.N. and J.L. prepared the samples. C.X. and X.M.X. grew the MnBi$_2$Te$_4$ bulk crystals. Y.P., X.C. and X.J. participated in the discussion. All authors discussed the results and contributed to the manuscript.

**Competing interests**

The authors declare no competing interests.

Supplementary Information for:

# Layer-dependent Magnetism with Even-Odd Effect and Magnetic Phase Diagrams of MnBi$_2$Te$_4$

Shiqi Yang[†], Xiaolong Xu[†] Yaozheng Zhu[†], Ruirui Niu, Chunqiang Xu, Yuxuan Peng, Xing Cheng, Xionghui Jia, Xiaofeng Xu, Jianming Lu[*] and Yu Ye[*]

[†]These authors contributed equally to this work

[*]Correspondence and request for materials should be addressed to J.L (jmlu@pku.edu.cn) and Y.Y. (email: ye_yu@pku.edu.cn)

**Contents:**

**S1. Reflective magnetic circular dichroism (RMCD) spectroscopy experimental setup**

**S2. Temperature-dependent Raman spectra of thickMnBi$_2$Te$_4$ and the low-frequency Raman spectroscopy experimental setup**

**S3. Atomic force microscopy (AFM) measurements of exfoliated MnBi$_2$Te$_4$ flakes**

**S4. Additional RMCD data for 1 SL sample under 532 nm excitation and 633 nm excitation**

**S5. Antiferromagnetic linear-chain model**

**S6. RMCD intensity maps including 3 SLs to 5 SLs**

**S7. Temperature-dependent RMCD measurements**

**S8. Magnetic state evolution at finite temperature predicted by a mean-field (MF) method**

# S1. Reflective magnetic circular dichroism (RMCD) spectroscopy experimental setup

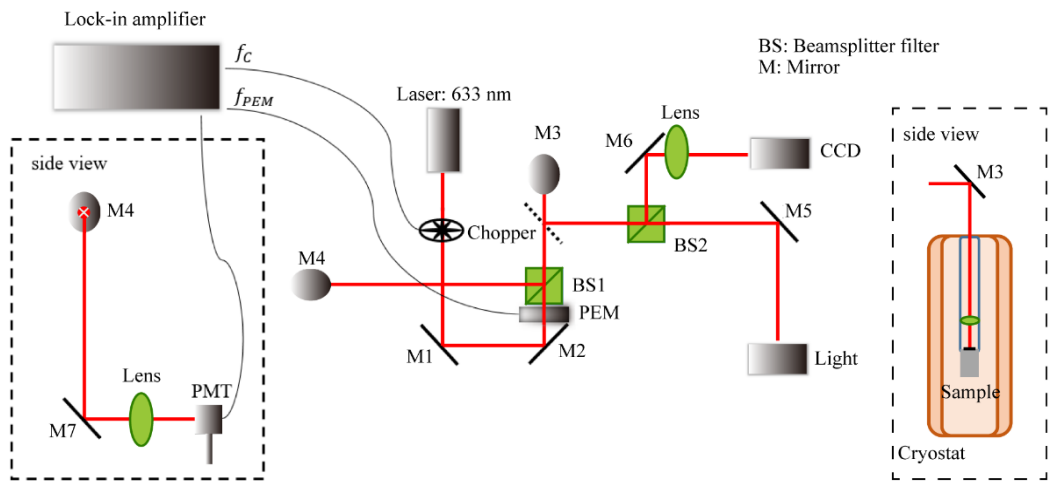

**Supplementary Figure 1.** RMCD spectroscopy experimental setup. Schematic of the optical setup used to measure the RMCD signals in the $MnBi_2Te_4$ samples. The He-Ne 633 nm laser is used as the excitation. A chopper and a photoelastic modulator (PEM) are used to modulate the intensity and polarization of the excitation beam, respectively. A magnetic field perpendicular to the sample plane is applied in the Attocube closed-cycle cryostat. The reflected beam was collected by the photomultiplier tube and was eventually analyzed by a two-channel lock-in amplifier through the signals of the reflected intensity (at $f_C$) and the RMCD intensities (at $f_{PEM}$).

## S2. Temperature-dependent Raman spectra of the thick MnBi$_2$Te$_4$ flakes

Temperature-dependent Raman spectroscopy was utilized to explore the temperature-dependent lattice structure transition or spin-phonon coupling. Supplementary Fig. 2 shows the schematic of the optical setup used to measure the temperature-dependent Raman spectra in the MnBi$_2$Te$_4$ samples. Resultant Raman signals were detected using an Andor spectrometer (SR-500i-D2-R) equipped with a Newton CCD (DU920P-BEX2-DD) with 1200 g/mm grating. The four observed Raman signatures, E$_g$ (47 cm$^{-1}$), A$_{1g}$ (66 cm$^{-1}$), E$_g^2$ (104 cm$^{-1}$), and A$_{1g}^2$ (139 cm$^{-1}$) shown in Supplementary Fig. 3 are independent with temperature, indicating no lattice structure transition in the cooling process.

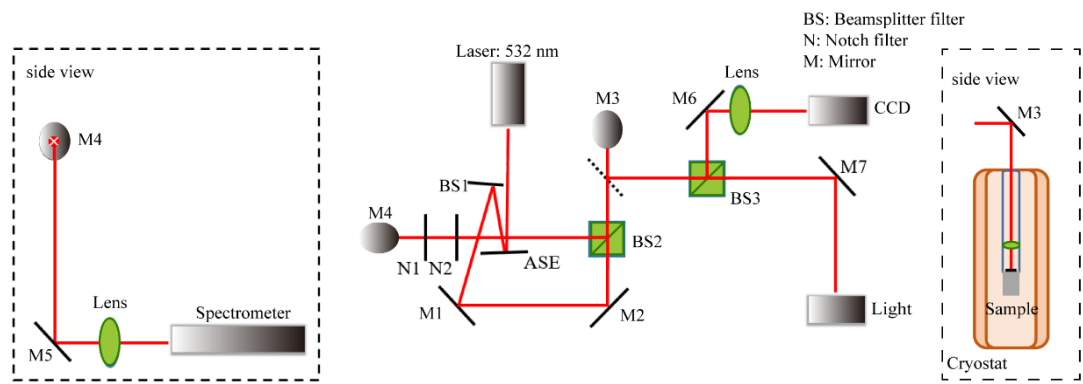

**Supplementary Figure 2.** Low-frequency Raman spectroscopy experimental setup. Schematic of the optical setup used to measure the low-frequency Raman in MnBi$_2$Te$_4$ samples. 532 nm laser was used as the excitation. ASE filter was used to suppress the broad spectrum of the spontaneous emission of the 532 nm laser. Two notch filters were used to filter the Rayleigh scattering to obtain the low-frequency Raman signals.

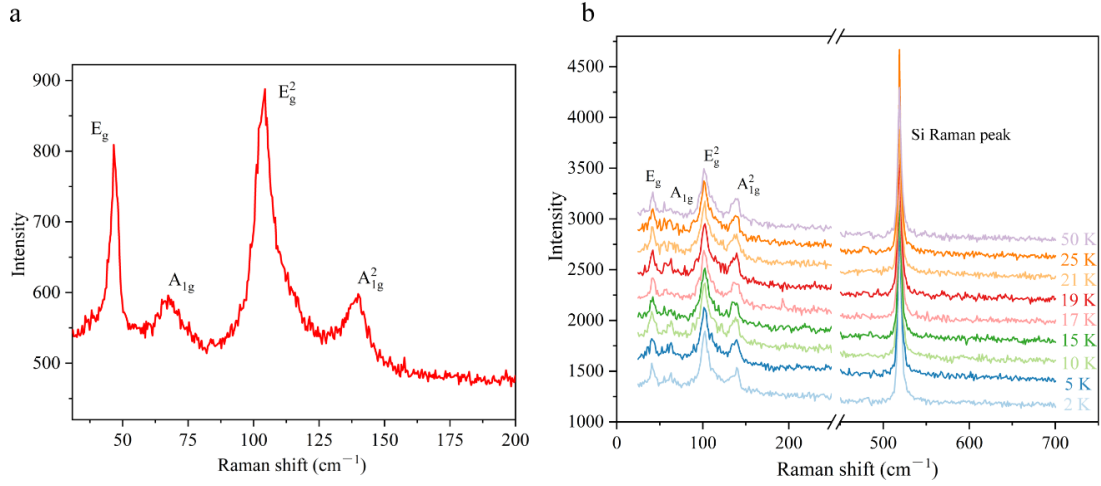

**Supplementary Figure 3.** Temperature-dependent Raman spectra. **a** Raman spectrum of thick MnBi$_2$Te$_4$ at room temperature, which is confirmed by the well-resolved E$_g$ (47 cm$^{-1}$), A$_{1g}$ (66 cm$^{-1}$), E$^2_g$ (104 cm$^{-1}$), and A$^2_{1g}$ (139 cm$^{-1}$) Raman signatures. **b** Temperature-dependent Raman spectra at a temperature range that passes through its $T_N$. The four Raman signatures show indistinguishable differences in our experimental configuration. The silicon's Raman signature at 520 cm$^{-1}$ was used as the reference.

# S3. Atomic force microscopy (AFM) measurements of the exfoliated MnBi$_2$Te$_4$ flakes

Before the AFM measurements, the PMMA covered on the MnBi$_2$Te$_4$ was removed using acetone, and then the sample was thoroughly rinsed with isopropyl alcohol (IPA). The detailed AFM images and height profiles of the sample measured in the main text are shown in Supplementary Fig. 4.

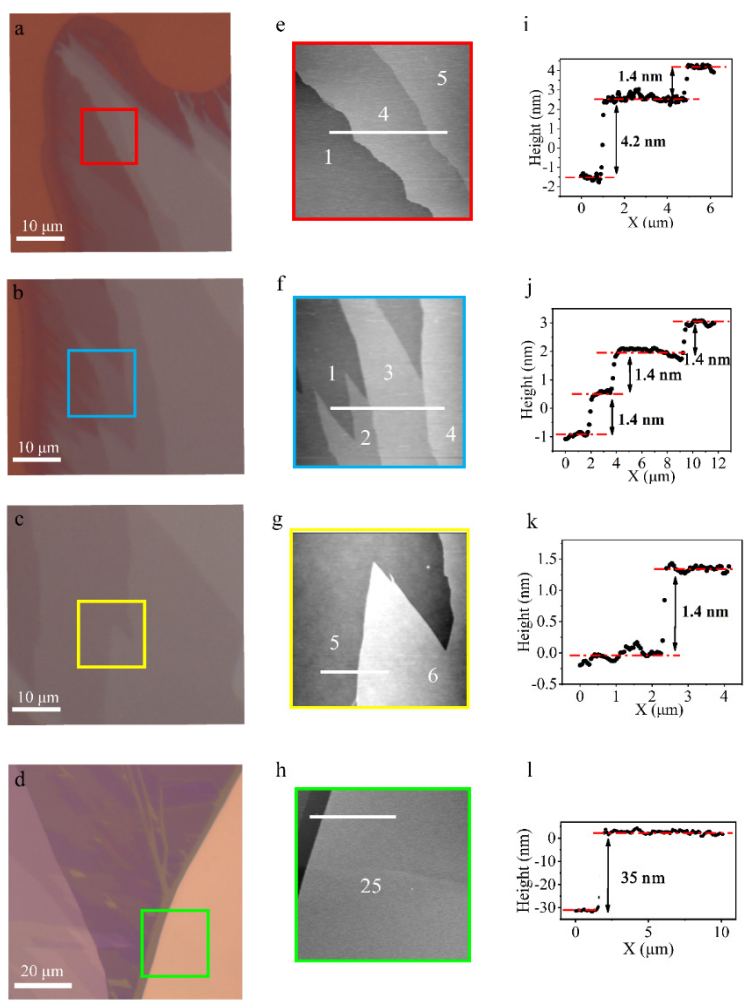

**Supplementary Figure 4. a-d** Optical images of the typical exfoliated few-$N$ SLs MnBi$_2$Te$_4$ flakes. The colored boxes denote the area for the corresponding AFM measurements obtained below. **e-h** AFM images of each area marked by corresponding boxes in optical images of **a-d**. **i-l** Line height profiles along the white lines marked in each map, indicating stepped MnBi$_2$Te$_4$ thicknesses from 2 SLs to 9 SLs (combing AFM data from the main text) and a ~25 SLs flake on the Au substrate.

**S4. RMCD data for 1 SL sample under 532 nm and 633 nm excitation**

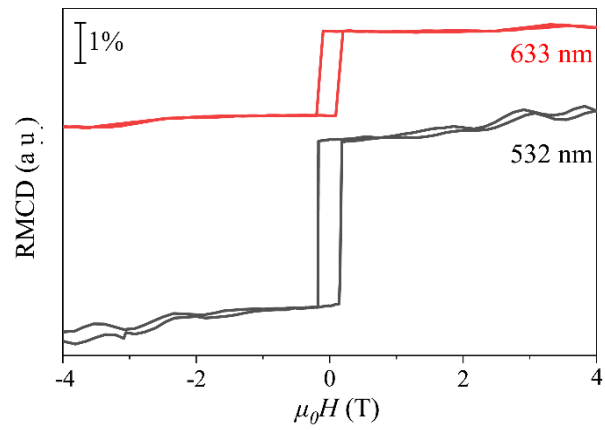

**Supplementary Figure 5.** RMCD data for 1 SL sample under 532 nm and 633 nm excitation. RMCD signals as a function the applied magnetic field for a 1 SL sample under 532 nm and 633 nm excitation at 1.6 K, showing identical critical field.

## S5. Antiferromagnetic linear-chain model

### Part I. Description of the model and its basic properties

As discussed in the main text, at the zero-temperature limit, we can assume a uniform magnetization within each layer and write the magnetization in the $i$-th layer as $\vec{M}_i = M_s(\sin\phi_i, 0, \cos\phi_i)$. For an $N$-layer system, the magnetic energy, corresponding to Eq. (1) in the main text, is

$$U_N = \mu_0 M_s \left[ \frac{H_J}{2} \sum_{i=1}^{N-1} \cos(\phi_i - \phi_{i+1}) - \frac{H_K}{2} \sum_{i=1}^{N} (\cos\phi_i)^2 - H \sum_{i=1}^{N} \cos\phi_i \right] \quad (1)$$

Solving the cases with small and soluble $N$ is of great help for us to illustrate the magnetic properties of this system. For a fixed $N$, to study the magnetic properties of the system, the following two conditions are critical:

➢ The ground state at each magnetic field $H$: This state is unique for each $H$, and can be obtained by finding the minimum energy given in Supplementary Eq. (1).

➢ The locally stable state(s) at each magnetic field $H$: For one specific $H$, there may exist more than one such states.

The ground state is useful in determining the general shape of the magnetization curve, while other locally stable states are essential in determining whether there is a hysteresis loop.

To find these states, we first find all the extreme points of this energy function, namely $U_N(\phi_1, \phi_2, \cdots, \phi_N)$, which are in general given by $\frac{\partial U_N}{\partial \phi_i} = 0$. This equation is equivalent to the following set of equations:

$$\begin{aligned}
&-\frac{H_J}{2}\sin(\phi_1 - \phi_2) + H_K\cos\phi_1\sin\phi_1 + H\sin\phi_1 = 0 \\
&-\frac{H_J}{2}\sin(\phi_2 - \phi_1) - \frac{H_J}{2}\sin(\phi_2 - \phi_3) + H_K\cos\phi_2\sin\phi_2 + H\sin\phi_2 = 0 \\
&\cdots \\
&-\frac{H_J}{2}\sin(\phi_i - \phi_{i-1}) - \frac{H_J}{2}\sin(\phi_i - \phi_{i+1}) + H_K\cos\phi_i\sin\phi_i + H\sin\phi_i = 0 \quad (2)\\
&\cdots \\
&-\frac{H_J}{2}\sin(\phi_{N-1} - \phi_{N-2}) - \frac{H_J}{2}\sin(\phi_{N-1} - \phi_N) + H_K\cos\phi_{N-1}\sin\phi_{N-1} + H\sin\phi_{N-1} = 0 \\
&-\frac{H_J}{2}\sin(\phi_N - \phi_{N-1}) + H_K\cos\phi_N\sin\phi_N + H\sin\phi_N = 0
\end{aligned}$$

Then, the local minimum states (denoted by $\phi_i^s$) are obtained, by filtering the extreme points under the condition that the matrix of second derivatives of $U_N$ is positive definite

$$\lambda_k(A^s) > 0, \text{ for all } k \ (A^s)_{ij} = \left[\frac{\partial^2 U_N}{\partial \phi_i \partial \phi_j}\right]_{\phi_i^s} \tag{3}$$

where $\lambda_k(A^s)$ are the eigenvalues of $A^s$. Then, the global minimum (ground state) is determined by finding the minimal $U_N$ within these solutions, which is easy because we only need to search in a finite set. It is straightforward to see that the solutions to Supplementary Eq. (2) contain some equivalent ones, for instance, the ones before and after the change $\phi_i \rightarrow -\phi_i$. To simplify the analysis, in the subsequent analysis, we usually take only one solution among the equivalent solutions.

**Part II. The magnetic state evolution in different anisotropy regions**

Three regions, namely low-anisotropy region (represented by $H_K/H_J = 0$), mid-anisotropy region (represented by $H_K/H_J = 0.3$) and high-anisotropy region (represented by $H_K/H_J = 0.6$) are explored in $N = 2$ and $N = 3$ systems. Both the total magnetization (along the $z$-axis) $M$ and the angles $\phi_i$ are shown in Supplementary Fig. 6, where $M$ is defined as

$$M = M_s \sum_{i=1}^{N} \cos\phi_i \tag{4}$$

There are three basic phases in this model: antiferromagnetic (AFM), canting AFM (CAFM) and ferromagnetic (FM). Here, $H_1$ is the critical field where the ground state changes away from the AFM phase (to either FM or CAFM), and $H_2$ is the critical field where the FM phase becomes locally unstable. In different regions ($H_K/H_J$ values), this model shows different properties, such as the behaviors in phase transitions, which are illustrated as follows. The role of anisotropy is of significant importance when considering these properties, which is not carefully covered in the previous studies[1].

➢ Low-anisotropy region: When $H_K/H_J = 0$, no hysteresis loop is observed, indicating a coincidence between the locally stable state and the ground state. For even $N$, $H_1$ is zero, indicating a spin-flop transition (from AFM to CAFM) at the zero field, while $H_1$ is finite for odd $N$. A spin-flip transition (from CAFM to FM) always happens at a finite magnetic field. There is no sudden change in $M$ at both $H_1$ and $H_2$. This coincides with the results in CrCl$_3$[1](Note that in the case of CrCl$_3$, the magnetic field is in the *x-y* plane and $\phi_i$ is the angle with the magnetic field).

- ➢ High-anisotropy region: A large hysteresis loop is dominated, with $H_2$ even smaller than $H_1$, and therefore the spin-flop transition cannot be observed. The absence of spin-flop transition is confirmed in the experiments related to CrI$_3$[3,4].

- ➢ Mid-anisotropy region: Here a small hysteresis loop occurs near $H_1$, shown by the difference between $H_\alpha$ and $H_\beta$. The spin-flop field $H_1$ and the spin-flip field $H_2$ are both finite, and a sudden change of magnetization happens at $H_1$. Since the hysteresis loop is relatively small, and the barrier between two locally stable states (when $H_\alpha < H < H_\beta$) is relatively low, it is reasonable to assume that only the ground state is dominant, and no loop or only a small loop (also observed in the previous works[5,6] for MnBi$_2$Te$_4$) can be observed near $H_1$.

All results show that this simple model works surprisingly well for all three regions.

At the end of this part, we describe an alternative method for calculating the evolution of magnetic states. It is noticeable that the most time-consuming step in the above algorithm is to find all the solutions to a set of equations because most numerical methods for solving equations or finding the minimum value rely on the given initial state to find a solution. To take advantage of such methods, we change the magnetic field from a very high magnitude (when all the magnetic moments are aligned in the same direction) step by step to zero and then return to the high magnitude step by step. In each step, we search the local minimum value of the energy function (1) in the $N$-dimensional space $\{\phi_i\}$, where the minimum value found at the previous step is used for initialization. This ensures that the magnetization and the magnetic state change continuously, unless at a critical magnetic field where a state is no longer locally stable. Thus, the outer loops in the $M - H$ curves can be easily obtained by this method.

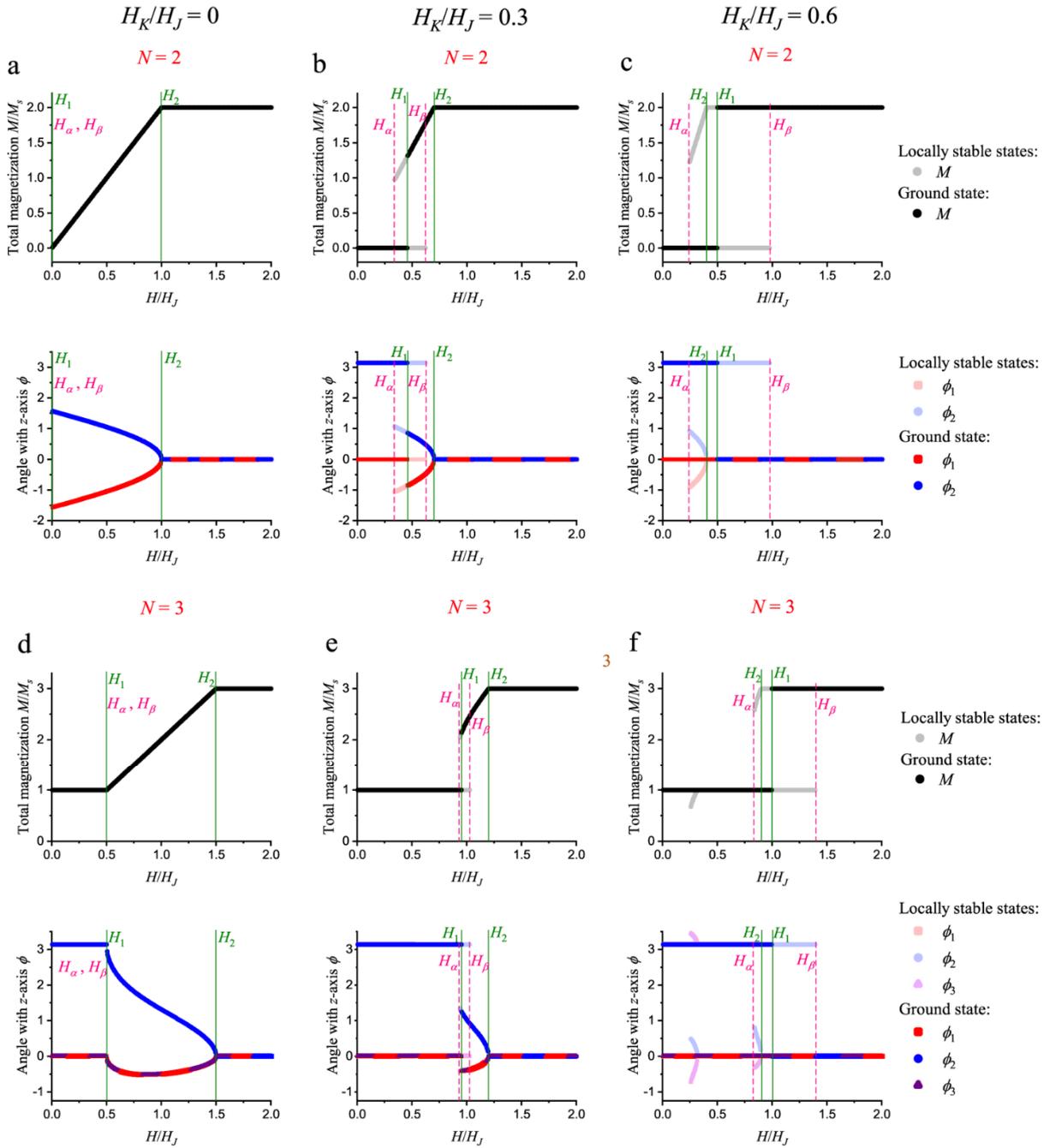

**Supplementary Figure 6.** Magnetic state evolutions for $N = 2$ and $N = 3$ as a function of the external field $H$ with different $H_K/H_J$ values. **a, d** Total magnetization along $z$-axis and angle with $z$-axis versus magnetic field ($M - H$ and $\phi_i - H$ diagrams) in low-anisotropy region with $H_K/H_J = 0$ for $N = 2$ (**a**) and $N = 3$ (**d**). The ground states are represented by dark black lines, and the locally stable states are represented by light gray lines in $M - H$ diagrams. Different colors in $\phi_i - H$ diagrams indicate different layers. $H_\alpha$ and $H_\beta$ (marked by pink dashed lines) denote the two critical magnetic fields for locally stable states (the boundaries of the possible hysteresis

loops), $H_1$ is the critical field where the ground state (dark lines) changes away from the AFM phase, and $H_2$ is the critical field where the FM phase becomes unstable ($H_1$ and $H_2$ are marked by green lines). **b, e** $M - H$ and $\phi_i - H$ diagrams in mid-anisotropy region with $H_K/H_J = 0.3$ for $N = 2$ (**b**) and $N = 3$ (**e**). **c, f** $M - H$ and $\phi_i - H$ diagrams in high-anisotropy region with $H_K/H_J = 0.6$ for $N = 2$ (**c**) and $N = 3$ (**f**).

## Part III. The spin-flop field in mid-anisotropy region

From now on, we consider the mid-anisotropy region, which is the case for MnBi$_2$Te$_4$, and only consider the ground state at each magnetic field. Finally, we will give the fitting results for $H_J$ and $H_K$.

At the exact spin-flop field $H = H_1$, there is a solution to $\phi_i$ (other than the AFM solution itself) to Supplementary Eq. (2), while also meets

$$U_N(\phi_1, \phi_2, \cdots, \phi_N) = U_N(0, \pi, 0, \pi, \cdots) \tag{5}$$

where the right side represents the energy of the AFM state. Refer to Supplementary Eq. (1), this additional condition further reads

$$\frac{H_J}{2}\sum_{i=1}^{N-1}\cos(\phi_i - \phi_{i+1}) - \frac{H_K}{2}\sum_{i=1}^{N}(\cos\phi_i)^2 - H\sum_{i=1}^{N}\cos\phi_i =$$

$$\begin{cases} \frac{N-1}{2}H_J - \frac{N}{2}H_K & \text{for even } N \\ \frac{N-1}{2}H_J - \frac{N}{2}H_K - H & \text{for odd } N \end{cases} \tag{6}$$

Thus, the spin-flop field $H_1$ can be directly obtained by numerically solving the Supplementary Eq. (2) and Eq. (6), treating $H$ as an unknown quantity. Sometimes more than one solution can be obtained, where the lowest positive solution $H$ represents the observed spin-flop field.

Now we can obtain the spin-flop field $H_1$ as a function of $N$, namely $H_1(N; H_J, H_K)$, where $H_J$ and $H_K$ are fitting parameters. These theoretical results are then compared with the experimental results, denoted as $H_1^{exp}(N)$. Then, we perform a standard $\chi^2$-fitting, that is, minimizing the quantity by fitting $H_J$ and $H_K$ appropriately.

$$\chi^2(H_J, H_K) = \sum_N \left[\frac{H_1(N; H_J, H_K) - H_1^{exp}(N)}{\Delta H_1^{exp}(N)}\right]^2 \tag{7}$$

Based on our experimental data, the summation over $N$ in Supplementary Eq. (7) from $N = 2$ to $N = 9$, and the uncertainty $\Delta H_1^{exp}(N)$ for each $N$ is also estimated from our experimental data.

Given the above fitting method, for MnBi$_2$Te$_4$, this model yields $\mu_0 H_J$ = 5.10 T and $\mu_0 H_K$ = 1.58 T, respectively. The results agree well with those of the first-principles calculations and other experiments reported elsewhere[2,7].

**Part IV. The magnetic state evolution of MnBi$_2$Te$_4$**

To further compare with our experimental results, it is very important to calculate the total magnetization (along the $z$-axis) $M$ defined in Supplementary Eq. (4) as a function of the external magnetic field $H$, as shown in Supplementary Fig. 7. The corresponding theoretical and experimental results of the spin-flop field $H_1$ (when the ground state changes from AFM to other states) are summarized in Fig. 2b and 2e in the main text. The fitting results are carefully analyzed in the main text, and our model describes the spin-flop field $\mu_0 H_1$ of different $N$ well.

Multi-step spin-flop transitions (from $H_1$ to $H_1'$) are observed in our calculation for even $N$ with $N \geqslant 4$, as shown in Supplementary Fig. 7. Supplementary Fig. 8 shows how the angle $\phi_i$ varies with the external field $H$ obtained from our calculation. For $N = 5$ (odd SLs), the ground state becomes symmetric ($\phi_1 = \phi_5$, $\phi_2 = \phi_4$) right after the spin-flop field $H_1$, while for $N = 4$ and 6 (even SLs), another transition (at $H_1'$) is needed to reach the symmetric state (e.g. $\phi_1 = -\phi_4$, $\phi_2 = -\phi_3$ for $N = 4$). Moreover, in our calculations, for $N = 6$ and 8, the slopes of the $M - H$ and the $\phi_i - H$ curves become very large shortly after $H_1'$, which indicates that a sharp coherent spin rotation occurs in a narrow magnetic field range (but still continuous).

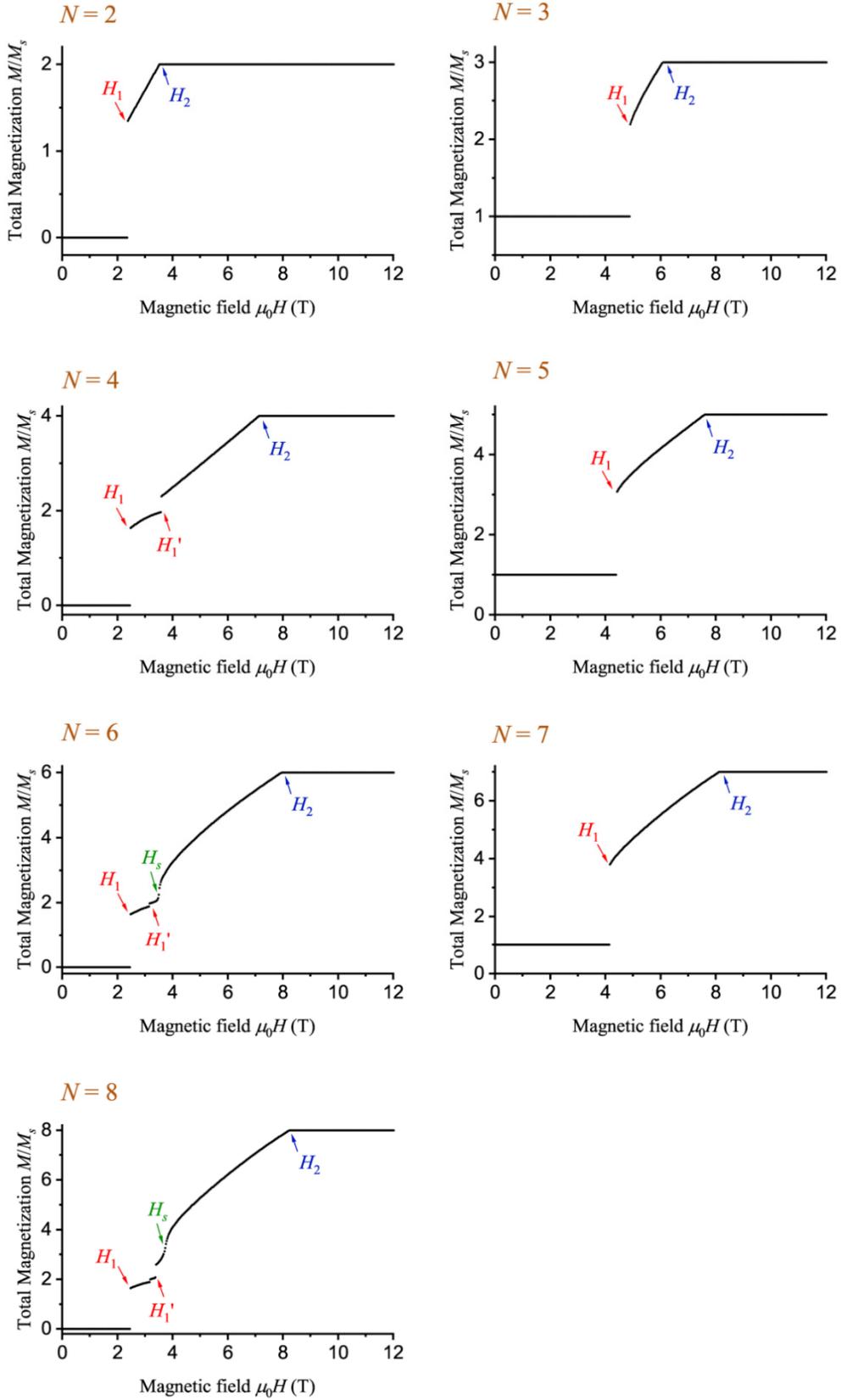

**Supplementary Figure 7.** The total magnetization along the $z$-axis (defined as $M = M_s \sum_{i=1}^{N} \cos\phi_i$) as a function of the external field $\mu_0 H$ for $N = 2 \sim 8$ of MnBi$_2$Te$_4$.

Only the ground states are shown. The spin-flop field $H_1$ for each $N$ is marked by the red arrow and the spin-flip fields $H_2$ by the blue arrow. For $N$ = 4, 6, and 8, the spin-flop transition experiences a multi-step process, and $H_1'$ represents the final spin-flop step. For $N$ = 6 and 8, a sharp coherent spin rotation in a narrow magnetic field range (at $H_s$, represented by the large slope of the curve and denoted by the green arrows) is also observed shortly after the final spin-flop process, which contributes to the second transition (see Fig. 2a in the main text) observed in our experiments.

A second transition is observed experimentally at 3.39±0.17 T for $N$ = 6 and 3.60± 0.13 T for $N$ = 8, respectively (indicated by the grey arrow in Fig. 2a in the main text), which may originate from the final step spin-flop transition at $H_1'$ (3.2 T for $N$ = 6 and 3.4 T for $N$ = 8), or the subsequent sharp coherent spin rotation in the narrow field range at $H_s$ (presented by the sudden increase of slope, 3.5 T for $N$ = 6 and 3.7 T for $N$ = 8, respectively). We note that our theory predicts a multiple-step spin-flop process not only for $N$ = 6 and 8 but for $N$ = 4, while a similar transition is not found experimentally in 4 SLs. In addition, the change of magnetization is remarkably small at $H_1'$, especially for $N$ = 6 (see Supplementary Fig. 7 for details). Therefore, we attribute the second transition for $N$ = 6 and 8 observed in the experiments to the subsequent sharp coherent spin rotation around $H_s$.

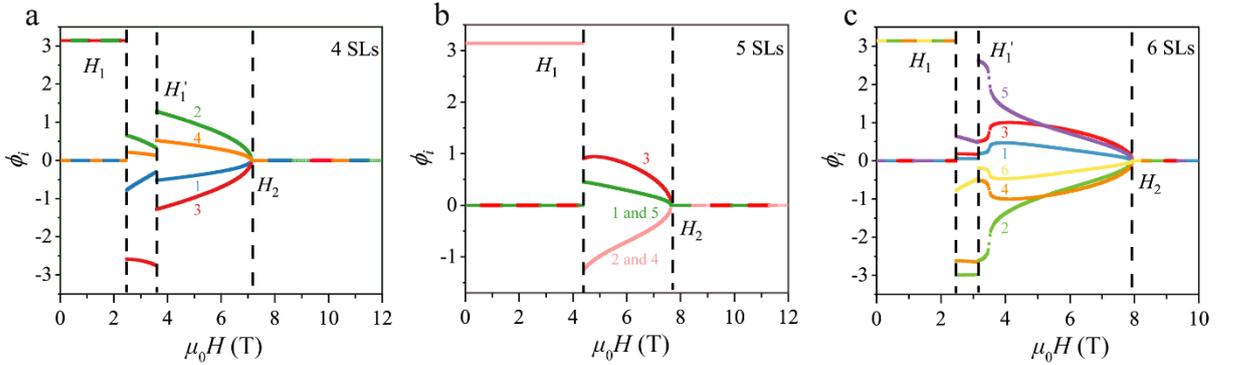

**Supplementary Figure 8.** The magnetic state evolution (angles $\phi_i$ of each layer) as a function of the external field $H$, by using the parameters $\mu_0 H_J$ = 5.14 T and $\mu_0 H_K$ = 1.58 T in 4 SLs (**a**) and 5 SLs (**b**) and 6 SLs (**c**) MnBi$_2$Te$_4$. In 4 SLs, the spin-flop transition occurs at ~ 2.5 T with a two-step process, and the canting evolves until the critical spin-flip field $H_2$ is reached. In 5 SLs, the antiferromagnetic state is stable up to ~ 4.5 T, at which a spin-flop transition occurs, then progressively canting until perfect alignment with the external field is reached at the spin-flip transition at $H_2$.

For 6 SLs, the spin-flop transition occurs at ~ 2.5 T with a two-step process. Different colors denote the evolution of different layers and numbers in figures represent layer numbers of the sample from top to bottom.

**Part V. Discussion about the spin-flip field $\mu_0 H_2$**

The spin-flip field, defined as the minimal magnetic field at which the FM state is locally stable, which can be derived from Supplementary Eq. (3) for the state $\phi_1 = \phi_2 = \cdots = \phi_N = 0$. For this state, the matrix of the second derivative of $U_N$, denoted as $A^{\text{FM}}$, is tridiagonal and can be explicitly calculated as

$$A^{\text{FM}} = \begin{pmatrix} H - \frac{H_J}{2} + H_K & \frac{H_J}{2} & & & & 0 \\ \frac{H_J}{2} & H - H_J + H_K & \frac{H_J}{2} & & & \\ & \frac{H_J}{2} & \ddots & \ddots & & \\ & & \ddots & \ddots & \frac{H_J}{2} & \\ & & & \frac{H_J}{2} & H - H_J + H_K & \frac{H_J}{2} \\ 0 & & & & \frac{H_J}{2} & H - \frac{H_J}{2} + H_K \end{pmatrix} \quad (8)$$

For a specific $N$, the eigenvalues of this matrix read

$$\lambda_k^{\text{FM}} = H - H_J + H_K + H_J \cos\left(\frac{k\pi}{N}\right), \quad k = 1, 2, \cdots, N-1 \quad (9)$$

Therefore, it is obvious that at a sufficiently high field, all eigenvalues $\lambda_k^{\text{FM}}$ are positive, and thus $A^{\text{FM}}$ is positive definite. Since the minimum eigenvalue corresponds to $k = N - 1$, it is straightforward that the spin-flip field $H_2$ is given by:

$$H_2 = 2H_J \cos^2\left(\frac{\pi}{2N}\right) - H_K \quad (10)$$

where $\lambda_{N-1}^{\text{FM}}$ vanishes.

For $N = 2$ and $N = 3$, this result is consistent with our experimental results, which is also verified in the main text. However, for $N \geqslant 4$, the spin-flip field is so large that it exceeds the magnetic field range we can apply experimentally. Moreover, due to the smooth transition process at $H_2$, the spin-flip field is hard to be accurately determined experimentally. Hence, $H_2$ is not used in the fitting of $H_J$ and $H_K$.

## S6. RMCD intensity maps of 3 SLs to 5 SLs samples

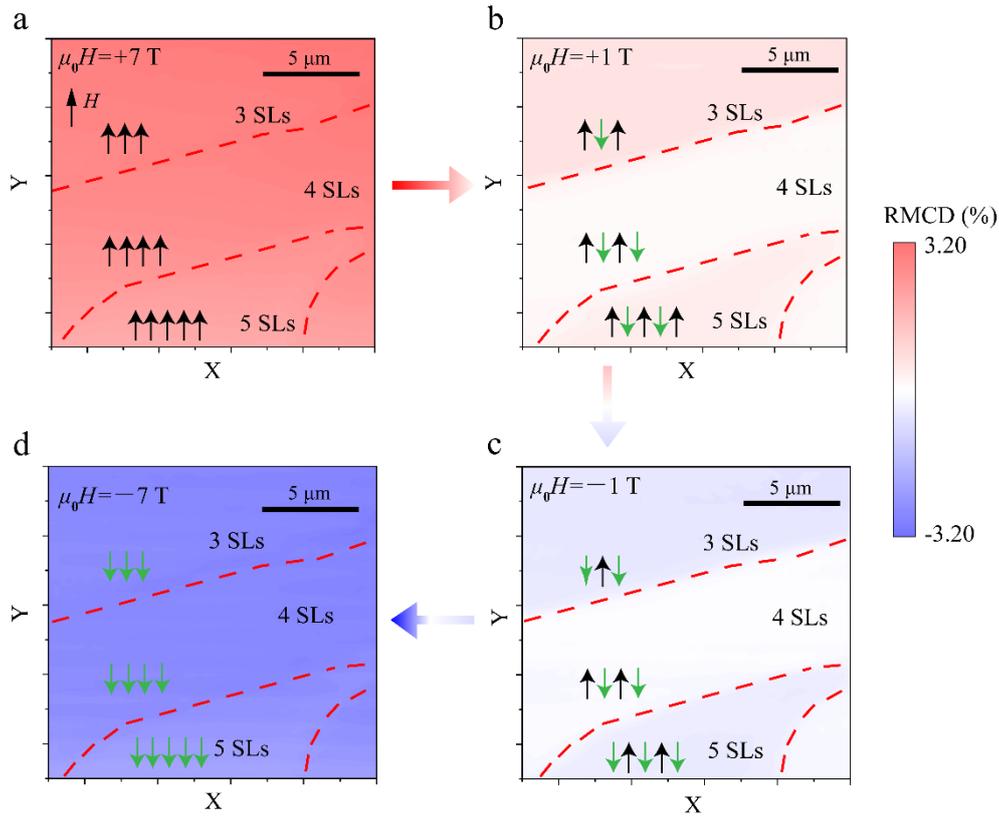

**Supplementary Figure 9.** Field-dependent RMCD intensity maps of 3 SLs to 5 SLs $MnBi_2Te_4$ at 2 K. **a** RMCD intensity maps at $\mu_0H = +7$ T on the stepped flake. The color from blue to red indicates the RMCD signal from negative to positive in three regions, linked to its magnetization. The inset arrows denote the spin orientation of each SL by the black (↑) and green (↓) colors in 3SLs, 4 SLs, and 5 SLs. **b** RMCD intensity maps at $\mu_0H = +1$ T on the stepped flake. **c** RMCD intensity maps at $\mu_0H = -1$ T on the stepped flake. The signals are symmetrical with those at $\mu_0H = +1$ T, and the magnetization is the opposite. **d** RMCD intensity maps at $\mu_0H = -7$ T on the stepped flake. The signals are symmetrical with those at $\mu_0H = +7$ T, and the magnetization is the opposite.

## S7. Temperature-dependent RMCD measurements

In addition to the thicknesses mentioned in the main text (1 SL, 2 SLs, 3 SLs, and 25 SLs), we also measured the temperature-dependent RMCD signals of 4 SLs, 5 SLs, and 6 SLs $MnBi_2Te_4$. As the temperature increases, the spin-flop field becomes smaller and eventually disappears at a specific temperature (that is how we define $T_N$ in even-$N$ samples).

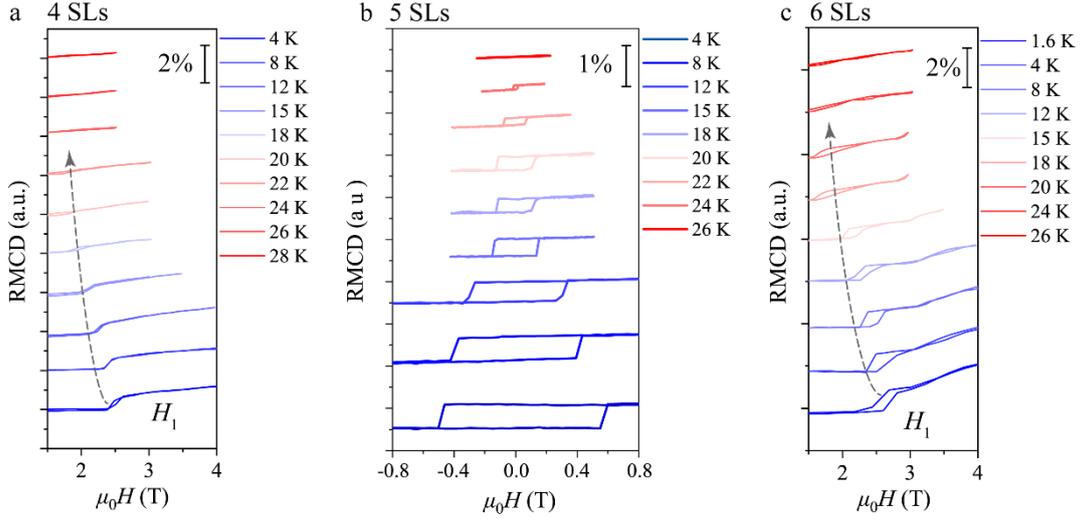

**Supplementary Figure 10.** Zoomed in temperature-dependent RMCD measurements of 4 SLs to 6 SLs. **a, c** The temperature-dependent RMCD measurements of 4 SLs and 6 SLs $MnBi_2Te_4$, showing a spin-flop transition at ~ 2.5 T for low-temperature, consistent with other even-$N$ samples. As the temperature increases, the spin-flop field becomes smaller and eventually disappears at a specific temperature ($T_N$). **b** The temperature-dependent RMCD measurements of 5 SLs $MnBi_2Te_4$.

## S8. Magnetic state evolution at finite temperature predicted by the mean-field method

### Part I. Description of the model

In the linear chain model described in Supplementary Information S5, only the ground state is considered, corresponding to zero temperature. The approach of using a representative spin to represent the magnetization in the entire layer is indeed strictly hold at zero temperature, while at a finite temperature requires a more complicated energy expression. By denoting the magnetization at position $a$ in the $i$-th layer as $\vec{M}_{i,a}$, the whole magnetic energy of an $N$-layer system under an external magnetic field $\vec{H}$ reads

$$U_N = \sum_{i=1}^{N-1}\sum_{a,b} J_{a,b} \frac{\vec{M}_{i,a} \cdot \vec{M}_{i+1,b}}{M_s^2} + \frac{1}{2}\sum_{i=1}^{N}\sum_{a,b} J_{a,b}^{\parallel} \frac{\vec{M}_{i,a} \cdot \vec{M}_{i,b}}{M_s^2} - \frac{K}{2}\sum_{i=1}^{N}\sum_{a}\left(\frac{\vec{M}_{i,a} \cdot \hat{z}}{M_s}\right)^2$$
$$-\mu_0 \vec{H} \cdot \sum_{i=1}^{N}\sum_{a} \vec{M}_{i,a} \qquad (11)$$

where $J_{a,b}$ stands for the interlayer interactions (only the interaction between the atoms in the nearest neighbor layers are accounted), $J_{a,b}^{\parallel}$ stands for the intralayer interactions and $K$ denotes the magnetic anisotropy energy. In general, $J_{a,b} = J_{b,a}$ and $J_{a,b}^{\parallel} = J_{b,a}^{\parallel}$.

Here, instead of completely solving the problem of this thermodynamics system, we perform a mean-field (MF) approximation to the intralayer interactions, which is effective if the temperature is not too high because the interlayer interactions are far smaller than the intralayer interacrtions[2]. In the finite-temperature linear chain model[1], only the ground state of the interlayer interactions is considered, and the layer magnetization at any temperature is required to reach definite predictions. Unlike the finite-temperature linear chain, this MF method takes into account the contribution of excited states and allows us to fully understand the system with a few predetermined parameters. (But sometimes, the excited states separated by a high barrier are not considered, resulting in the hysteresis loop described in Part III.) Under this approximation, the coupling terms between different sites disappear, and Supplementary Eq. (11) becomes

$$U_N^{\text{MF}} = \sum_{i=1}^{N-1}\sum_{a,b} J_{a,b} \frac{\langle\vec{M}_{i,a}\rangle \cdot \vec{M}_{i+1,b}}{M_s^2} + \sum_{i=1}^{N-1}\sum_{a,b} J_{a,b} \frac{\vec{M}_{i,a} \cdot \langle\vec{M}_{i+1,b}\rangle}{M_s^2} - \sum_{i=1}^{N-1}\sum_{a,b} J_{a,b} \frac{\langle\vec{M}_{i,a}\rangle \cdot \langle\vec{M}_{i+1,b}\rangle}{M_s^2}$$
$$+ \sum_{i=1}^{N}\sum_{a,b} J_{a,b}^{\|} \frac{\vec{M}_{i,a} \cdot \langle\vec{M}_{i,b}\rangle}{M_s^2} - \frac{1}{2}\sum_{i=1}^{N}\sum_{a,b} J_{a,b}^{\|} \frac{\langle\vec{M}_{i,a}\rangle \cdot \langle\vec{M}_{i,b}\rangle}{M_s^2} - \frac{K}{2}\sum_{i=1}^{N}\sum_{a}\left(\frac{\vec{M}_{i,a} \cdot \hat{z}}{M_s}\right)^2$$
$$-\mu_0 \vec{H} \cdot \sum_{i=1}^{N}\sum_{a} \vec{M}_{i,a} \qquad (12)$$

which is known as the MF energy. After "decoupling" the spins at different sites, we can take a representative spin for each layer and write the corresponding *N*-moment energy as:

$$U_N^{\text{MF},1}(\vec{M}_1, \vec{M}_2, \ldots, \vec{M}_N) = \sum_{i=1}^{N-1} J \frac{\langle\vec{M}_i\rangle \cdot \vec{M}_{i+1} + \vec{M}_i \cdot \langle\vec{M}_{i+1}\rangle}{M_s^2} + \sum_{i=1}^{N} J^{\|} \frac{\vec{M}_i \cdot \langle\vec{M}_i\rangle}{M_s^2} - \frac{K}{2}\sum_{i=1}^{N}\left(\frac{\vec{M}_i \cdot \hat{z}}{M_s}\right)^2$$
$$-\mu_0 \vec{H} \cdot \sum_{i=1}^{N} \vec{M}_i - \sum_{i=1}^{N-1} J \frac{\langle\vec{M}_i\rangle \cdot \langle\vec{M}_{i+1}\rangle}{M_s^2} - \frac{1}{2}\sum_{i=1}^{N} J^{\|} \frac{\langle\vec{M}_i\rangle \cdot \langle\vec{M}_i\rangle}{M_s^2} \qquad (13)$$

where

$$J = \sum_a J_{ab} = \sum_b J_{ab}, \quad J^{\|} = \sum_a J_{a,b}^{\|} = \sum_b J_{a,b}^{\|} \qquad (14)$$

and the total MF energy Supplementary Eq. (12) is then expressed as

$$U_N^{\text{MF}} = \sum_a U_N^{\text{MF},1}(\vec{M}_{1,a}, \vec{M}_{2,a}, \ldots, \vec{M}_{N,a}) \qquad (15)$$

The parameters of this MF model take the same values as the fitting results of the linear chain model at zero-temperature, i.e. $\mu_0 H_J = (2J)/M_s = 5.10$ T and $\mu_0 H_K = K/M_s = 1.58$ T. The intralayer interaction (favorable for ferromagnetic alignment) is set to be $\mu_0 H_J^{\|} = (2J^{\|})/M_s = -54.0$ T, which is a much larger than the interlayer interactions.

**Part II. The equivalence with the linear chain model at zero temperature**

To show the equivalence between this model and the linear chain model (discussed in detail in Supplementary Information S5) at zero temperature, it is worth noting that the statistical average value defined in Eq. (7) in the main text will collapse to the ground state of the energy Supplementary Eq. (13). At the same time, the statistical average value of $\langle\vec{M}_i\rangle$ is exactly $M_s$ at zero temperature, but usually smaller than $M_s$ at a finite temperature. Note that the second term in Supplementary Eq. (13) (intralayer interaction) takes its minimum value at the exact point $\vec{M}_i = \langle \vec{M}_i \rangle$, and the self-consistent condition also requires the minimum value at the point of $\vec{M}_i = \langle \vec{M}_i \rangle$ for the rest parts

$$U_N^{\text{MF},1,\star} = \sum_{i=1}^{N-1} J \frac{\langle \vec{M}_i \rangle \cdot \vec{M}_{i+1} + \vec{M}_i \cdot \langle \vec{M}_{i+1} \rangle}{M_s^2} - \frac{K}{2} \sum_{i=1}^{N} \left( \frac{\vec{M}_i \cdot \hat{z}}{M_s} \right)^2 - \mu_0 \vec{H} \cdot \sum_{i=1}^{N} \vec{M}_i \quad (16)$$

After using a similar formulism in Supplementary Information S5 Part I, this energy function will yield the same result as Supplementary Eq. (2), showing that these two models are in general the same at zero temperature.

**Part III. The temperature – field phase diagram**

For a fixed layer number $N$ and temperature $T$, the magnetization-field curve ($M - H$ curve) can be calculated by using the method described in Methods in the main text, and then the total magnetization (along the $z$-direction) is expressed as:

$$M = \sum_{i=1}^{N} \langle \vec{M}_i \rangle_z \quad (17)$$

The magnetic field $\mu_0 H$ changes from 12 T to zero and then returns to 12 T in steps of 0.05 T. The $M - H$ curve from zero temperature up to over 24 K is summarized to draw the phase diagram, in which temperature interval is as small as 0.5 K. Using such methods, the field-temperature phase diagrams for $N = 2$ and $N = 3$ are shown in Fig. 4 in the main text, and phase diagrams for $N = 4$, 5, and 6 are shown in Supplementary Fig. 11. The two local minima from the up-sweeping and down-sweeping, which result in the hysteresis loop, are judged by averaging the MF energy in Supplementary Eq. (13) for all accounted statuses, and only the one with smaller energy is taken. This avoids the complicated hysteresis loop problem and returns to the previous definition of the linear chain model at zero temperature (the ground state at zero temperature, see Supplementary Information S5). However, for a finite temperature, in the hysteresis loop, the two states correspond to two locally minimums, and the occupancy rates of two local minimums cannot be fully determined by their minimum energy. Thus, the less occupied minimum can also have important effects and this technique is an estimation.

With the complicated hysteresis loops removed, we can define the spin-flop and spin-flip fields. At a certain temperature, when the magnetic field is small enough, the state was AFM. As the magnetic field increases, the magnetization gradually increases, corresponding to a non-zero susceptibility, until it jumps to another state (CAFM state) with far larger magnetization at the spin-flop field $H_1$. Then, the magnetization evolves rapidly until reaching the FM state, leaving only the paramagnetic background (see $M - H$ curves in Supplementary Fig. 12 at temperatures near $T_N$). At the spin-flip field $H_2$, a slope change in the $M - H$ curve is observed at the boundary between CAFM and FM states. For even-$N$ samples with $N \geqslant 4$, the spin-flop transition undergoes a multi-step process. The final spin-flop transition occurs at the critical field $H_1'$, in which the ground state changes from an asymmetric state to a symmetric state (see Supplementary Information S5 Part IV for details). These phase boundaries (shown in Fig. 4 in the main text and Supplementary Fig. 11 by dotted lines) are determined by connecting the critical fields calculated at different temperatures, and smoothing is applied to eliminate the undulation from the random nature of our algorithm. Then the three phase regions (AFM, CAFM, and FM) for all layer numbers are determined.

The success of our temperature-dependent model is reflected by the consistency of the theoretical predictions with our experimental data (up to at least 18 K for $N = 2$ and 22 K for $N \geqslant 3$). As the temperature approaches the Néel temperature, the theoretical prediction gradually deviates from the experimental data, because fluctuations of two-dimensional magnetic systems become more dominant in this region, which is ignored in the MF method. For example, for $N > 3$, the Néel temperature $T_N$ predicted by the MF model is about 30% higher than the experimental value of ~25 K. All in all, it is surprising that our MF model describes the system very well below the Néel temperature, with only three parameters $H_J, H_K$ (taken from the fitting result in Supplementary Information S5 Part III) and $H_J^\parallel$.

The theoretical values of the final spin-flop step at $H_1'$ and sharp coherent spin rotation at $H_s$ of $\langle \vec{M}_i \rangle_z - H$ curves are plotted in blue and red in Supplementary Fig. 11.

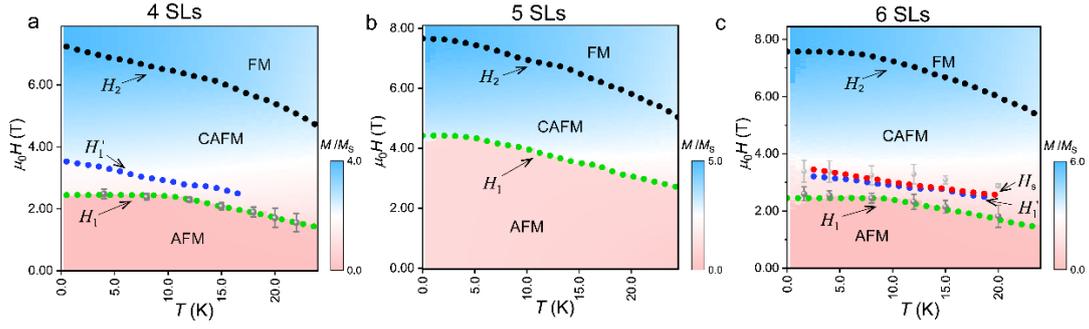

**Supplementary Figure 11.** Temperature – field phase diagrams calculated using the MF methods, for $N = 4$ (**a**), $N = 5$ (**b**) and $N = 6$ (**c**). The green, blue, and black dotted lines represent the calculated spin-flop field $H_1$, $H_1'$ and spin-flip field $H_2$ (smoothed), respectively. The red dotted line represents the field $H_s$ where a sharp coherent spin rotation occurs. The experimental data are represented in grey ($H_1$) and light grey ($H_s$) circles with corresponding error bars.

In addition, when we study the $M - H$ curves (with loops, see Supplementary Fig. 12) at different temperatures, some of the characteristics in the experimental curves can be explained qualitatively:

- ➢ As the temperature increases, the hysteresis loop near the spin-flop field $H_1$ (pointed by black arrows in Supplementary Figure 12) becomes smaller and eventually disappears, and the spin-flop transition changes from a direct and sharp jump to a smooth change process in the $M - H$ curves. In 3 SLs, at 26 K, it is hard to distinguish the spin-flop transition from the magnetization changes in the CAFM region. In our experiments, such behavior is also found near $T_N$, resulting in remarkably large errors in the critical fields measured in this region.

- ➢ At very low temperatures, the slope change at the spin-flip field $H_2$ (marked by cyan arrows in Supplementary Figure 12) is very sharp, and a clear discontinuity of the second derivative is seen in the curves. However, at higher temperatures, the slope change becomes remarkably smoother, and it turns out that it is difficult to determine the spin-flip field $H_2$ at high temperatures. For higher temperatures, the paramagnetic signal in the FM region is also larger. These effects explain why $H_2$ is shallowed by the paramagnetic background and cannot be accurately obtained in our measurements at temperatures above ~ 14 K.

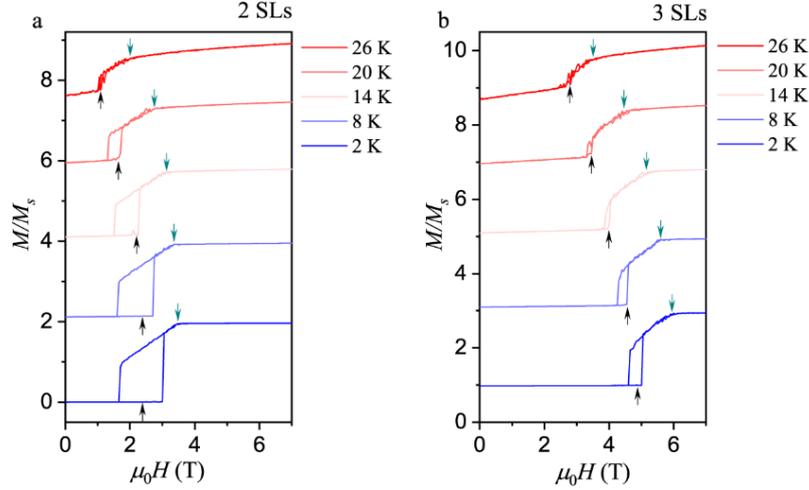

**Supplementary Figure 12.** Temperature-dependent $M – H$ curves calculated using the MF methods, for $N = 2$ (**a**) and $N = 3$ (**b**). Different colors represent different temperatures, and the curves are offset for clarity. For each curve, the spin-flop field $H_1$ is marked by a black arrow, and the spin-flip field $H_2$ is marked by a cyan arrow.

In summary, our MF model agrees quantitively well with all of our experiment data, as long as the temperature is below the Néel temperature, and helps us understand many characteristics in the experimental curves qualitatively. However, due to the experimental limitations, the verification of the spin-flip field is hard to perform quantitively, where a larger magnetic field and more precise techniques are needed. Additionally, to understand the behavior of this system around the Néel temperature, other theoretical models (such as Monte Carlo simulations[2] directly employed to the original energy Supplementary Eq. (11)), might be applied.